\documentclass[twocolumn,times]{aastex63}

\expandafter\def\csname editcolor1\endcsname{magenta}
\expandafter\def\csname editcolor2\endcsname{blue}  
\expandafter\def\csname editcolor3\endcsname{violet} 

\usepackage{newtxtext,newtxmath}
\usepackage[T1]{fontenc}
\usepackage{graphicx}	
\usepackage{amsmath}	
\usepackage{float}
\usepackage{bookmark} 
\bookmarksetup{numbered, open,}
\usepackage{enumitem}
\setlist[enumerate]{itemsep=0mm}
\usepackage{hyperref}
\usepackage{subfigure}
\usepackage{xspace}
\usepackage{xcolor} 

\newcommand{\eg}{e.g.\xspace}
\newcommand{\ie}{i.e.\xspace}
\newcommand{\Mch}{\ensuremath{M_\text{Ch}}\xspace}
\newcommand{\Msun}{\ensuremath{M_{\odot}}\xspace}

\newcommand{\kms}{km~s\ensuremath{^{-1}}\xspace}

\newcommand{\gcm}{g cm\ensuremath{^{-3}}\xspace}

\newcommand{\Nifs}{\ensuremath{^{56}}Ni\xspace}

\newcommand{\mic}{\ensuremath{\mu}m\xspace}
\newcommand{\DmB}{\Delta{\rm m_{15}(B)}\xspace}

\graphicspath{{./}{figs/}}

\received{January 9, 2023}
\revised{February, 1, 2023}
\accepted{February 1, 2023}

\submitjournal{ApJL}

\shorttitle{SN~2021aefx: High-density Burning in SNe~Ia}
\shortauthors{DerKacy, Ashall, Hoeflich, Baron et al.}

\begin{document}
\title{JWST Low-Resolution MIRI Spectral Observations of SN~2021aefx: High-density Burning in a Type Ia Supernova}

\correspondingauthor{James M DerKacy}
\email{jmderkacy@vt.edu}

\author[0000-0002-7566-6080]{J.~M.~DerKacy}
\affiliation{Department of Physics, Virginia Tech, Blacksburg, VA 24061, USA}

\author[0000-0002-5221-7557]{C.~Ashall}
\affiliation{Department of Physics, Virginia Tech, Blacksburg, VA 24061, USA}

\author[0000-0002-4338-6586]{P.~Hoeflich}
\affiliation{Department of Physics, Florida State University, 77 Chieftan Way, Tallahassee, FL 32306, USA}

\author[0000-0001-5393-1608]{E.~Baron}
\affiliation{Homer L. Dodge Department of Physics and Astronomy, University of Oklahoma, 440 W. Brooks, Rm 100, Norman, OK 73019-2061, USA}
\affiliation{Hamburger Sternwarte, Gojenbergsweg 112, D-21029 Hamburg, Germany}

\author[0000-0003-4631-1149]{B.~J.~Shappee}
\affiliation{Institute for Astronomy, University of Hawai'i at Manoa, 2680 Woodlawn Dr., Hawai'i, HI 96822, USA }

\author[0000-0003-1637-9679]{D.~Baade}
\affiliation{European Organization for Astronomical Research in the Southern Hemisphere (ESO), Karl-Schwarzschild-Str. 2, 85748 Garching b. M\"unchen, Germany}

\author[0000-0003-0123-0062]{J.~Andrews}
\affiliation{Gemini Observatory/NSF’s NOIRLab, 670 North A`ohoku Place, Hilo, HI 96720-2700, USA}

\author[0000-0002-4924-444X]{K.~A.~Bostroem}
\altaffiliation{LSSTC Catalyst Fellow}
\affiliation{Steward Observatory, University of Arizona, 933 North Cherry Avenue, Tucson, AZ 85721-0065, USA}

\author[0000-0001-6272-5507]{P.~J.~Brown}
\affiliation{George P. and Cynthia Woods Mitchell Institute for Fundamental Physics and Astronomy, Texas A\&M University, Department of Physics and Astronomy, College Station, TX 77843, USA}

\author[0000-0003-4625-6629]{C.~R.~Burns}
\affiliation{Observatories of the Carnegie Institution for Science, 813 Santa Barbara Street, Pasadena, CA 91101, USA}

\author[0000-0002-5380-0816]{A.~Burrow}
\affiliation{Homer L. Dodge Department of Physics and Astronomy, University of Oklahoma, 440 W. Brooks, Rm 100, Norman, OK 73019-2061, USA}

\author[0000-0001-7101-9831]{A.~Cikota}
\affiliation{Gemini Observatory/NSF's NOIRLab, Casilla 603, La Serena, Chile}

\author[0000-0001-6069-1139]{T.~de~Jaeger}
\affiliation{Institute for Astronomy, University of Hawai'i at Manoa, 2680 Woodlawn Dr., Hawai'i, HI 96822, USA }

\author[0000-0003-3429-7845]{A.~Do}
\affiliation{Institute for Astronomy, University of Hawai'i at Manoa, 2680 Woodlawn Dr., Hawai'i, HI 96822, USA }

\author[0000-0002-7937-6371]{Y.~Dong}
\affiliation{Department of Physics, University of California, 1 Shields Avenue, Davis, CA 95616-5270, USA}

\author[0000-0002-3827-4731]{I. Dominguez}
\affiliation{Universidad de Granada, 18071, Granada, Spain}

\author[0000-0002-1296-6887]{L.~Galbany}
\affiliation{Institute of Space Sciences (ICE, CSIC), Campus UAB, Carrer de Can Magrans, s/n, E-08193 Barcelona, Spain}
\affiliation{Institut d’Estudis Espacials de Catalunya (IEEC), E-08034 Barcelona, Spain}

\author[0000-0003-1039-2928]{E.~Y.~Hsiao}
\affiliation{Department of Physics, Florida State University, 77 Chieftan Way, Tallahassee, FL 32306, USA}

\author[0000-0001-6209-838X]{E.~Karamehmetoglu}
\affiliation{Department of Physics and Astronomy, Aarhus University, Ny Munkegade 120, DK-8000 Aarhus C, Denmark}

\author[0000-0002-6650-694X]{K.~Krisciunas}
\affiliation{George P. and Cynthia Woods Mitchell Institute for Fundamental Physics and Astronomy, Texas A\&M University, Department of Physics and Astronomy, College Station, TX 77843, USA}

\author[0000-0001-8367-7591]{S.~Kumar}
\affiliation{Department of Physics, Florida State University, 77 Chieftan Way, Tallahassee, FL 32306, USA}

\author[0000-0002-3900-1452]{J.~Lu}
\affiliation{Department of Physics, Florida State University, 77 Chieftan Way, Tallahassee, FL 32306, USA}

\author[0000-0001-5888-2542]{T.~B.~Mera~Evans}
\affiliation{Department of Physics, Florida State University, 77 Chieftan Way, Tallahassee, FL 32306, USA}

\author[0000-0003-0733-7215]{J.~R.~Maund}
\affiliation{Department of Physics and Astronomy, University of Sheffield, Hicks Building, Hounsfield Road, Sheffield S3 7RH, U.K.}

\author[0000-0001-6876-8284]{P.~Mazzali}
\affiliation{Astrophysics Research Institute, Liverpool John Moores University, UK}
\affiliation{Max-Planck Institute for Astrophysics, Garching, Germany}

\author[0000-0001-7186-105X]{K.~Medler}
\affiliation{Astrophysics Research Institute, Liverpool John Moores University, UK}
\affiliation{Max-Planck Institute for Astrophysics, Garching, Germany}

\author[0000-0003-2535-3091]{N.~Morrell}
\affiliation{Las Campanas Observatory, Carnegie Observatories, Casilla 601, La Serena, Chile}
	
\author[0000-0002-0537-3573]{F.~Patat}
\affiliation{European Organization for Astronomical Research in the Southern Hemisphere (ESO), Karl-Schwarzschild-Str. 2, 85748 Garching b. M\"unchen, Germany}

\author[0000-0003-2734-0796]{M.~M.~Phillips}
\affiliation{Las Campanas Observatory, Carnegie Observatories, Casilla 601, La Serena, Chile}

\author[0000-0002-9301-5302]{M.~Shahbandeh}
\affiliation{Space Telescope Science Institute, 3700 San Martin Drive, Baltimore, MD 21218-2410, USA}

\author[0000-0001-5570-6666]{S.~Stangl}
\affiliation{Homer L. Dodge Department of Physics and Astronomy, University of Oklahoma, 440 W. Brooks, Rm 100, Norman, OK 73019-2061, USA}

\author[0000-0003-0763-6004]{C.~P.~Stevens}
\affiliation{Department of Physics, Virginia Tech, Blacksburg, VA 24061, USA}

\author[0000-0002-5571-1833]{M.~D.~Stritzinger}
\affiliation{Department of Physics and Astronomy, Aarhus University, Ny Munkegade 120, DK-8000 Aarhus C, Denmark}

\author[0000-0002-8102-181X]{N.~B.~Suntzeff}
\affiliation{George P. and Cynthia Woods Mitchell Institute for Fundamental Physics and Astronomy, Texas A\&M University, Department of Physics and Astronomy, College Station, TX 77843, USA}

\author[0000-0002-0036-9292]{C.~M.~Telesco}
\affiliation{Department of Astronomy, University of Florida, Gainesville, FL 32611 USA}

\author[0000-0002-2471-8442]{M.~A.~Tucker}
\altaffiliation{CCAPP Fellow}
\affiliation{Center for Cosmology and AstroParticle Physics, The Ohio State University, 191 W. Woodruff Ave., Columbus, OH 43210, USA}

\author[0000-0001-8818-0795]{S.~Valenti}
\affiliation{Department of Physics, University of California, 1 Shields Avenue, Davis, CA 95616-5270, USA}

\author[0000-0001-7092-9374]{L.~Wang}
\affiliation{Department of Physics and Astronomy, Texas A\&M University, College Station, TX 77843, USA}

\author[0000-0002-6535-8500]{Y.~Yang}
\altaffiliation{Bengier-Winslow-Robertson Postdoctoral Fellow}
\affiliation{Department of Astronomy, University of California, Berkeley, CA 94720-3411, USA}

\author[0000-0001-8738-6011]{S.~ W.~Jha}
\affiliation{Department of Physics and Astronomy, Rutgers, the State University of New Jersey, 136 Frelinghuysen Road, Piscataway, NJ 08854-8019, USA}

\author[0000-0003-3108-1328]{L.~A.~Kwok}
\affiliation{Department of Physics and Astronomy, Rutgers, the State University of New Jersey, 136 Frelinghuysen Road, Piscataway, NJ 08854-8019, USA}

\begin{abstract}
We present a JWST/MIRI low-resolution mid-infrared (MIR) spectroscopic
observation of the normal Type Ia supernova (SN Ia) SN~2021aefx at
$+323$~days past rest-frame $B$-band maximum light. The spectrum ranges
from 4-14~\mic, and shows many unique qualities including a
flat-topped [\ion{Ar}{3}]~$8.991$~\mic profile, a strongly tilted
[\ion{Co}{3}]~$11.888$~\mic feature, and multiple stable Ni lines. These
features provide critical information about the physics of the explosion.
The observations are compared to synthetic spectra from detailed NLTE
multi-dimensional models. The results of the best-fitting model are used
to identify the components of the spectral blends and provide a
quantitative comparison to the explosion physics. Emission line profiles
and the presence of electron capture (EC) elements are used to constrain
the mass of the exploding white dwarf~(WD) and the chemical asymmetries
in the ejecta. We show that the observations of SN~2021aefx are
consistent with an off-center delayed-detonation explosion of a
near-Chandrasekhar mass~(\Mch)~WD at a viewing angle of $-30\arcdeg$
relative to the point of the deflagration-to-detonation transition. From
the strength of the stable Ni lines we determine that there is little to
no mixing in the central regions of the ejecta. Based on both
the presence of stable Ni and the Ar velocity distributions, we obtain a
strict lower limit of $1.2$~\Msun the initial WD, implying that most
sub-\Mch explosions models are not viable models for SN~2021aefx. The
analysis here shows the crucial importance of MIR spectra for
distinguishing between explosion scenarios for SNe~Ia.
\end{abstract}



\section{Introduction} \label{sec:intro}

Type Ia supernovae (SNe~Ia) arise from the thermonuclear explosion of at 
least one carbon/oxygen (C/O) white dwarf (WD) in a binary system \citep{Hoyle1960}. 
Despite being the most precise extra-galactic distance indicators in the 
Universe 
\citep{Phillips1993,Riess1998,Perlmutter1999,Riess2016,Riess2018}, 
the exact make-up of SNe~Ia progenitor systems and the mechanism of their 
explosions are still unknown 
\citep[see][for recent reviews]{Maoz:2014,BranchWheeler2017,Jha2019}.

There are multiple progenitor scenarios that may produce SNe Ia. These 
include: the single-degenerate (SD) scenario where the companion is a 
main-sequence star or an evolved, non-degenerate companion like a red 
giant or He-star \citep{Whelan1973}; the double-degenerate (DD) scenario, 
where the companion is also a WD \citep{Iben1984,Webbink1984}; or a 
triple system where at least two of the bodies are C/O WDs 
\citep{Thompson2011,Kushnir2013}. Additionally, a wide range of explosion 
mechanisms also exist. Multiple mechanisms originate from the merger of
both stars in the progenitor system, including the dynamical merger of two WDs
\citep{Benz_etal_1990,Garcia-Berro_etal_2017}, the violent merger of 
two WDs \citep{Pakmor2012,Pakmor2013}, and the collisions of two WDs 
within triple systems \citep{Rosswog2009,Kushnir2013}. Currently, two
of the leading explosion models are of SNe Ia arising from the explosion
of a near-\Mch mass  WD, and the detonation of a 
sub-\Mch mass WD. In \Mch explosions H, He, or C 
material is accreted from a companion star (which can be degenerate 
or non-degenerate) until the central density in the primary WD is 
high enough  to trigger a thermonuclear 
runaway \citep{Iben1984,Diamond_etal_2018}. The flame can then propagate as 
a deflagration, detonation or both via a 
deflagration-to-detonation transition (DDT) 
\citep{Khokhlov91,Hoeflich_Khokhlov_1996,Gamezo03,Poludnenko19}. 
In contrast, a sub-\Mch explosion is triggered when a surface He 
layer detonates and drives a shock-wave inwards, causing a secondary 
detonation that disrupts the whole WD 
\citep{Nomoto84,Woosley94,Livne95,Hoeflich_Khokhlov_1996,Shen:2018}. 
Similar to a \Mch explosion, a sub-\Mch explosion can occur in both 
the single and double degenerate scenarios \citep{Piersanti03a,Piersanti03b}.

Due to the degenerate nature of C/O WDs, the central density~($\rho_c$) 
of the star is directly correlated with its mass \citep{Chandra_SS39}. 
Therefore, one of the key differences between \Mch and sub-\Mch scenarios 
is the peak density of the burning. In particular when 
$\rho_{c} > 5\times 10^8$~\gcm, 
significant amounts of stable iron group elements~(IGEs) such as 
$^{58}$Ni are produced \citep{Hoeflich_Khokhlov_1996}. These central densities 
correspond to WD masses of $ \gtrsim 1.2$~\Msun, where the thermonuclear 
runaway must start via compressional heating in the center of the WD 
\citep{Seitenzahl:2017}.   

Traditionally there have been fewer studies of SNe~Ia in the longer 
near-infrared~(NIR) and mid-infrared~(MIR) wavelengths compared to the 
optical. However, recent efforts have shown that these longer wavelengths 
offer additional, and sometimes better, information about the physics of 
SN explosions  
\citep{Meikle1993,Hoeflich_Gerardy_2002,Marion:2009,Hsiao_2013,Graham_etal_2017,Diamond_etal_2018,Wilk2018,Hoeflich_2021_20qxp,Lu:2023,2022arXiv221006993K,Hoeflich_etal_2023_19np}. 
This is due, in part, to the fact that the location of the photosphere is 
wavelength-dependent, and that different diagnostic spectral lines are 
revealed at longer wavelengths 
\citep{Hoeflich_1991,Hoeflich_1995,Wheeler1998,Kasen_2006,Ashall19a,Ashall19b}.

Prior to the launch of the James Webb Space Telescope (JWST), there 
were only seven MIR ($\lambda > 5$~\mic) spectral observations of 
SNe~Ia across four different objects. Three spectra were obtained 
with the Spitzer Space Telescope (SST); one of SN~2003hv at 
$\sim +375$~days (relative to estimated explosion), one of SN~2005df 
at $\sim +135$~days \citep{Gerardy2007}, and one of SN~2006ce at 
$+127$~days relative to $B$-band maximum light 
(\citealp[GO-30292, PI: W.P. Meikle;][]{Kwok2022}). Four 
MIR spectra of SN~2014J were obtained with CanariCam on the 10.4-m 
Gran Telescopio Canarias (GTC) between $57-137$~days after explosion 
\citep{Telesco_etal_2015}. Despite the small sample size it is apparent 
that the MIR contains many diagnostics to differentiate between leading 
explosion scenarios. For example, nebular phase MIR spectral observations, 
which probe the high-density central layers, can reveal the presence and 
distribution of stable Ni. These lines are direct indicators of 
high-density burning.

With the successful launch of JWST, high-S/N MIR spectral observations 
during the nebular phase of SNe Ia are now possible. The first spectrum 
of a SN~Ia obtained with JWST was that of SN~2021aefx at $+255$~days after 
maximum light (\citealp[MJD=59801.4;][]{Kwok2022}). Here we present and 
analyze a spectrum of SN~2021aefx taken $+323$~days (MJD=59871.6) after maximum light. In contrast to the work of \citet{Kwok2022}, who focused 
primarily on line identifications and determination of observed velocities, 
we interpret the explosion physics of SN~2021aefx through comparisons to a 
self-consistent set of non-local thermodynamic equilibrium~(NLTE) 
radiation hydrodynamic models of SNe Ia. This allows us to provide
a set of line IDs specific to SN~2021aefx in addition to a consistent 
picture of the explosion based on our newly observed spectrum and models.
In \autoref{sec:Data} we describe our observations, and in \autoref{sec:DR}
the details of our spectral reduction. Line identifications from full NLTE 
models are performed in \autoref{sec:IDs}, while an analysis of their 
velocities is presented in \autoref{sec:Vel}. \autoref{sec:Mod} discusses 
the details of our chosen NLTE models and a comparison to the observations. 
Alternative explosion scenarios are discussed in \autoref{sec:alt_scenario}. 
Finally, we summarize our findings in \autoref{sec:conclus}.

\section{Observations} \label{sec:Data}

SN~2021aefx was discovered on 2021 Nov 11.3 (MJD=59529.5) by the
Distance Less Than 40 Mpc Survey (DLT40; \citealp{DLT40})
and classified as a young SN~Ia \citep{Bostroem21,Hosseinzadeh22}. 
SN~2021aefx was subsequently followed by several groups, including 
a multi-band optical and spectroscopic follow-up campaign by the
Precision Observations of Infant Supernova Explosions Collaboration
(\href{https://poise.obs.carnegiescience.edu/}{POISE}, \citealp{Burns2021,Ashall22}). 
POISE's detailed photometric observations revealed an early blue 
excess, which may be explained by a rapid change in the velocities 
of spectral lines \citep{Ashall22}. 
An analysis of the complete POISE data set reveals the basic light 
curve properties of SN~2021aefx, including a decline rate of 
$\DmB=1.01 \pm 0.06$~mag, and a peak absolute magnitude of 
$M_{B} =-19.28 \pm 0.49$~mag, which places SN~2021aefx in the 
normal part of the luminosity-width relation 
(\citealp{Phillips1993,Ashall22}{; C. Stevens et al., in prep.}).
SN~2021aefx is located 105\arcsec.3 south, 37\arcsec.0 west from the 
center of its host NGC 1566 at a redshift of 0.005
($\alpha = 04^{h}20^{m}00^{s}.42$, $\delta=-54\arcdeg56\arcmin16\arcsec.10$; 
\citealp{Allison14}). 
NGC 1566 is a face-on spiral galaxy with systemic recessional velocity 
of 1500~\kms, and a rotational velocity of $65 \pm 60$~\kms at the location
of the SN \citep{Elagali2019}. All figures showing observed spectra of 
SN~2021aefx have been corrected for the combined recessional and rational
velocities of 1550~\kms at the location of the SN in the host.
This low rotational velocity implies that any observed off-center lines
(\ie lines shifted relative to the line-of-sight velocity) are intrinsic 
to the progenitor system itself, and not attributable to a peculiar 
velocity within the host galaxy.  

We present a MIR observation of SN~2021aefx obtained through program
GO-JWST-2114 (P.I. Ashall) from $\sim 4-14$~\mic. The 
data were obtained using JWST's Mid-Infrared Instrument (MIRI) in its 
Low Resolution Spectroscopy (LRS) configuration. In this mode, MIRI/LRS 
obtains slit spectroscopy of objects with a spectral resolving power 
$(R=\lambda/\Delta\lambda)$ of $R\sim100$ at $7.5$~\mic, varying 
from $R \sim 40$ at $5$~\mic to $R \sim 160$ at $10$~\mic
\citep{Kendrew15,Kendrew16,Rigby22}. The instrumental configuration is
identical to that of \citet{Kwok2022}. The spectral observations were 
performed with a 2-point dither strategy. For each grating setting 
there were $134$ groups per integration, $2$ integrations per exposure 
and $1$ exposure per dither. This results in an exposure of 
$734.5$~seconds at each dither position, which are combined for a 
total exposure time of $1493$~seconds. Full details of our observational
set-up are found in \autoref{tab:obs}.

\begin{figure*}
    \centering
    \includegraphics[width=\textwidth]{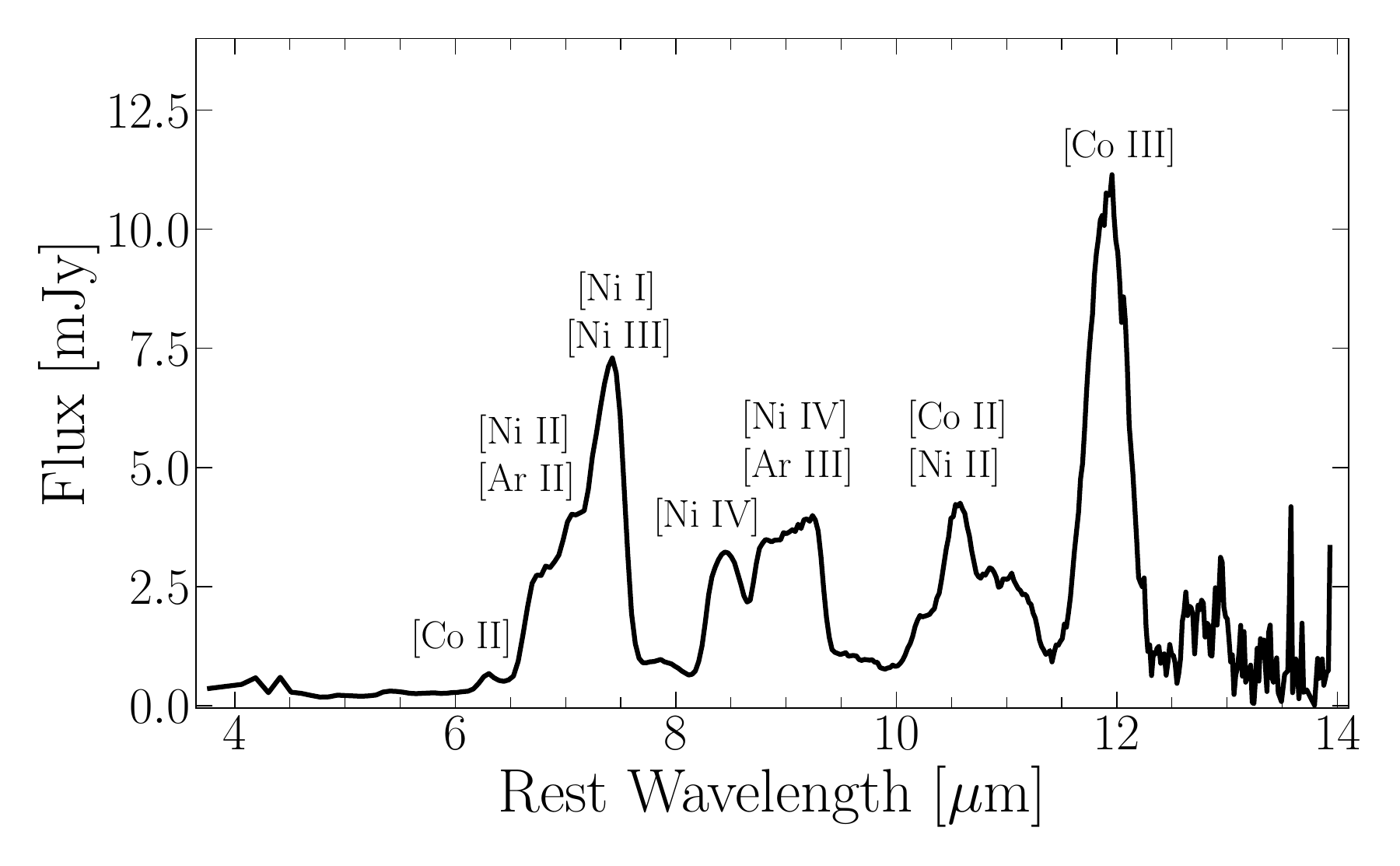}
    \caption{JWST/MIRI LRS spectrum of SN~2021aefx at $+323$ days 
    relative to $B$-band maximum. The ions responsible for the
    most prominent features in the spectrum are labeled. A full 
    set of line identifications is plotted in \autoref{fig:IDS}
    and shown in \autoref{tab:mir_lines}.}
    \label{fig:full_spec}
\end{figure*}

\begin{deluxetable}{cc}
  \tablecaption{JWST/MIRI Observation Details \label{tab:obs}} 
  \tablehead{\colhead{Parameter} & \colhead{Value}}
  \startdata
    Acquisition Image & \\
    \hline
    Filter & F1000W \\
    Exp Time [s] & 89 \\
    Readout Pattern & FASTGRPAVG8 \\
    \hline
    SN~2021aefx Spectrum & \\
    \hline
    Mode & LRS \\
    Exp Time [s] & 1493 \\
    $T_\text{obs}$ [MJD] & 59871.6 \\
    Epoch\tablenotemark{a} [days] & 322.71 \\ 
    Groups per Integration & 134 \\
    Integrations per Exp. & 2 \\
    Exposures per Dither & 1 \\
    Total Dithers & 2 \\
  \enddata
  \tablecomments{~\tablenotemark{a}~Rest frame days relative to 
  $B$-band maximum of MJD $=59547.25$ (C.~Stevens et al., in prep.).}
\end{deluxetable}

\section{Data Reduction} \label{sec:DR}

The data were obtained on 2022 Oct 19.6 (MJD=59871.6) and reduced with 
the JWST calibration
pipeline\footnote{\url{https://jwst-pipeline.readthedocs.io/en/stable/jwst/introduction.html}},
version 1.8.1 \citep{Bushouse2022_JWSTpipeline}.
Both raw (Stage 1 calibrated) and fully reduced data 
were retrieved from the Mikulski Archive for Space Telescopes 
(MAST)\footnote{\url{https://mast.stsci.edu/portal/Mashup/Clients/Mast/Portal.html}}.
This data can be accessed via: 
\dataset[DOI: 10.17909/6fjc-sx91]{https://doi.org/10.17909/6fjc-sx91}. 
The raw data was processed with a local installation of the 
version 1.8.1 pipeline for comparison to the fully reduced data from 
MAST, using the \texttt{spec\_mode\_stage\_2} and
\texttt{spec\_mode\_stage\_3} Jupyter notebooks as templates for the
reduction. Both reductions used the most up-to-date wavelength 
(jwst\_miri\_specwcs\_0005.fits) and flux calibration 
(jwst\_miri\_photom\_0085.fits) files. These calibration files
produce a wavelength solution accurate to $\sim0.05-0.02$~\mic,
varying from short to long wavelengths, and a flux calibration
accurate to a $\sim 2$ -- $5$\% global offset between 5 -- 12~\mic.
(\citealp[][S.~Kendrew, private communication]{Gordon2022}). Furthermore, 
\citet{Kwok2022} found that the flux calibration of their MIRI spectrum was 
accurate to 2\%.

Using the LRS Optimal Spectral Extraction 
notebook\footnote{\url{https://spacetelescope.github.io/jdat\_notebooks/notebooks/MIRI\_LRS\_spectral\_extraction/miri\_lrs\_spectral\_extraction.html}}, 
the spectra were re-extracted using multiple techniques. This re-extraction 
was necessary to properly center the position of the spectrum in the science 
aperture, as the pipeline-derived aperture produced a poor extraction at long
wavelengths. After proper re-extraction with the Optimal Extraction notebook, 
no significant differences were found between the locally reduced data and 
the fully calibrated (but un-extracted) data available from MAST. Future 
updates to the JWST calibration files are expected to further improve the 
accuracy of the automated extractions (S.~Kendrew, private communication). 

\begin{figure}
    \centering
    \includegraphics[width=\columnwidth]{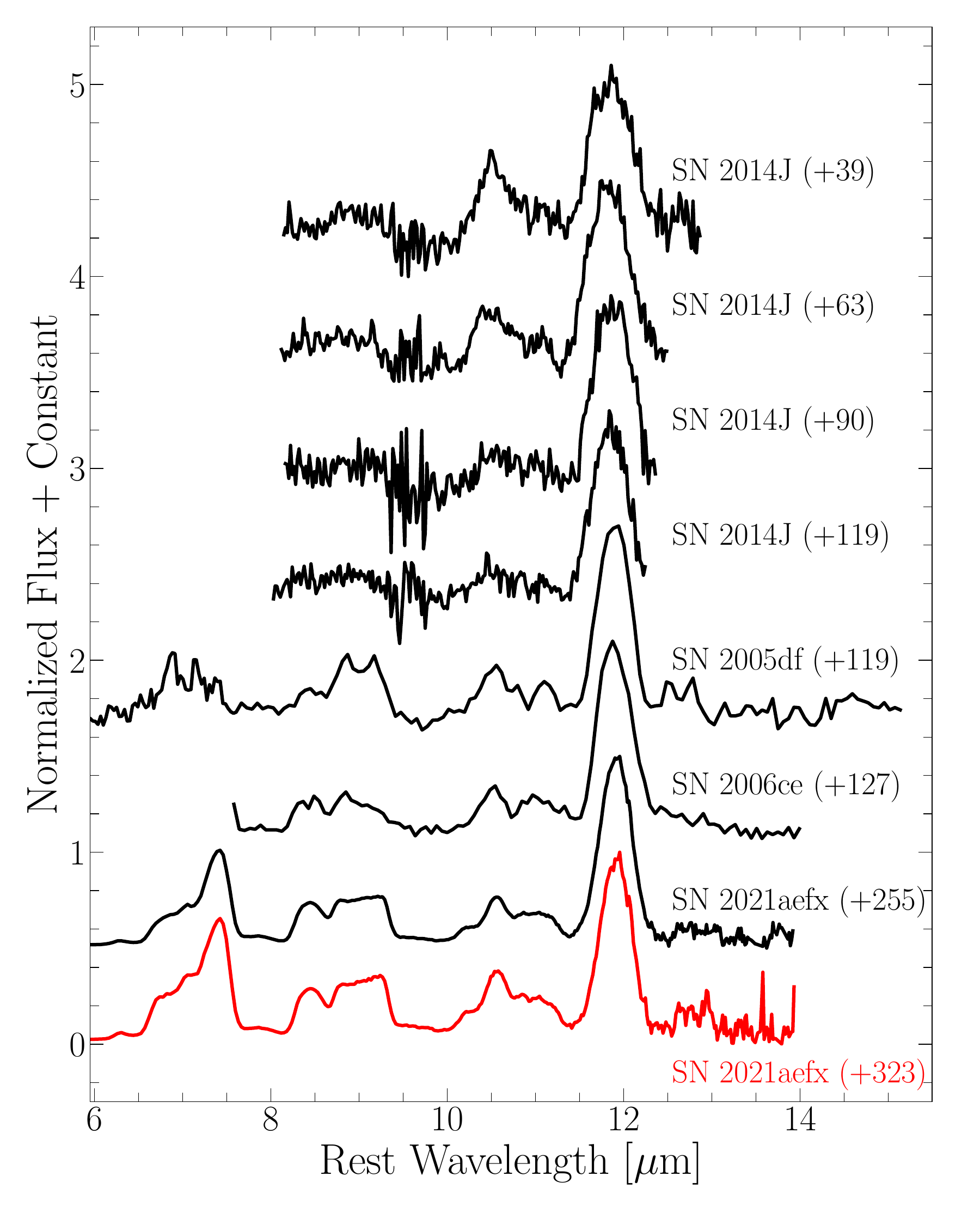}
    \caption{Comparison of $+323$~day spectrum of SN~2021aefx to 
    other MIR spectral observations of SNe Ia, including SNe~2005df 
    \citep{Gerardy2007}, 2006ce \citep{Kwok2022}, 2014J 
    \citep{Telesco_etal_2015}, and the $+255$ days spectrum of 
    2021aefx \citep{Kwok2022}. The epoch relative to $B$-band maximum 
    light for each spectrum is shown. The primary difference between the
    $+255$ and $+323$ day spectra of SN~2021aefx is the increased 
    strength of other features relative to the peak at $\sim 11.9$~\mic.}
    \label{fig:mir_compare}
\end{figure}

\section{Data comparison \& Line Identifications} \label{sec:IDs}

\autoref{fig:full_spec} presents the spectrum of SN~2021aefx 
acquired on 2022 Oct 19.6 (corresponding to $+323$~days after
$B$-band maximum light) from $4-14$~\mic. At these phases, the 
ejecta are optically thin and dominated by emission lines. The 
strongest of these lines are labeled in \autoref{fig:full_spec}.  
\autoref{fig:mir_compare} shows our spectrum of SN~2021aefx 
compared to the MIR spectra of SNe~2005df \citep{Gerardy2007}, 
2006ce \citep{Kwok2022}, 2014J \citep{Telesco_etal_2015}, and the 
earlier spectrum of SN~2021aefx at +$255$~days \citep{Kwok2022}. From 
\autoref{fig:mir_compare} it is clear that SN~2021aefx is similar to 
other previously observed SNe~Ia, but the size and sensitivity of 
JWST produces a high S/N spectrum with a quality that 
was previously impossible to obtain. Comparing the two JWST spectra of 
SN~2021aefx, the most noticeable difference is the decrease in the 
relative strength of the $\sim11.9$~\mic profile compared to the 
other features caused by the radioactive decay of $^{56}$Co.

To assist in line identifications, we use a suite of full NLTE
radiation transport models. These models reproduce both the early and
late time properties of SN~2021aefx, and an in-depth discussion of
the models with respect to the  MIR observables can be found in 
\autoref{sec:Mod}. 

\begin{figure*}
    \centering
    \includegraphics[width=\textwidth]{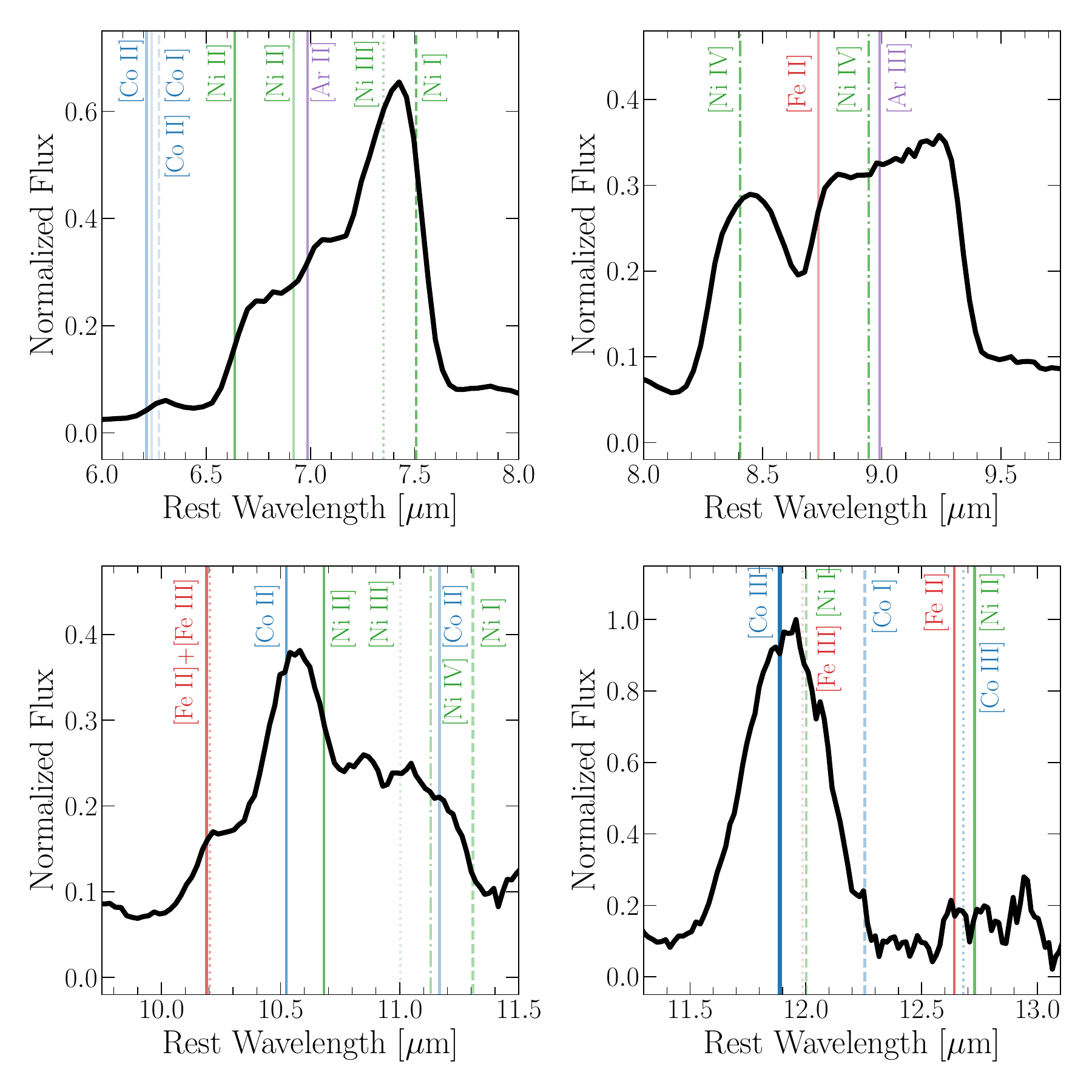}
    \caption{Detailed line identifications in the four prominent 
    feature regions based on the lines from Model 25 \citep{Hoeflich_2017} 
    included in \autoref{tab:mir_lines}. The color intensity of 
    the vertical lines corresponds to the strength of the spectral lines, 
    with 4-star lines the most intense and 1-star lines being the 
    faintest. Dashed lines correspond to ground state ions, solid lines 
    singly-ionized species, dotted lines doubly-ionized species, and 
    dash-dotted lines triply-ionized species.}
    \label{fig:IDS}
\end{figure*}

A detailed examination of the SN~2021aefx MIR spectrum reveals four prominent 
wavelength regions of line formation, which are described individually
in the following subsections. Detailed line identifications in each of these 
regions are plotted in \autoref{fig:IDS}, while \autoref{tab:mir_lines} 
lists the lines that contribute significantly to the spectrum.

\begin{deluxetable}{rcl|rcl}
  \tablecaption{Mid-Infrared Line Identifications from Model 25 \label{tab:mir_lines}}
  \tablehead{\colhead{\bf S} & \colhead{$\lambda$~[\mic]} & \colhead{Ion} & 
    \colhead{\bf S} & \colhead{$\lambda$~[\mic]} & \colhead{Ion}}
    \startdata
    $    *\ *$ & $6.214$ & [\ion{Co}{2}] & $          $ & $8.555$ & [\ion{Fe}{3}] \\
    $       *$ & $6.273$ & [\ion{Co}{1}] & $      *\ *$ & $8.611$ & [\ion{Fe}{3}] \\
    $       *$ & $6.274$ & [\ion{Co}{2}] & $         *$ & $8.644$ & [\ion{Co}{2}] \\
    $    *\ *$ & $6.383$ & [\ion{Ar}{3}] & $      *\ *$ & $8.733$ & [\ion{Fe}{2}] \\
    $ *\ *\ *$ & $6.636$ & [\ion{Ni}{2}] & $      *\ *$ & $8.945$ & [\ion{Ni}{4}] \\
    $        $ & $6.636$ & [\ion{Ni}{2}] & $   *\ *\ *$ & $8.991$ & [\ion{Ar}{3}] \\
    $    *\ *$ & $6.920$ & [\ion{Ni}{2}] & $         *$ & $9.618$ & [\ion{Ni}{2}] \\
    $ *\ *\ *$ & $6.985$ & [\ion{Ar}{2}] & $         *$ & $10.080$ & [\ion{Ni}{2}] \\ 
    $        $ & $6.985$ & [\ion{Ar}{2}] & $   *\ *\ *$ & $10.189$ & [\ion{Fe}{2}] \\
    $        $ & $6.985$ & [\ion{Ar}{2}] & $      *\ *$ & $10.203$ & [\ion{Fe}{3}] \\
    $        $ & $7.045$ & [\ion{Co}{1}] & $   *\ *\ *$ & $10.523$ & [\ion{Co}{2}] \\
    $        $ & $7.103$ & [\ion{Co}{3}] & $   *\ *\ *$ & $10.682$ & [\ion{Ni}{2}] \\
    $        $ & $7.147$ & [\ion{Fe}{3}] & $         *$ & $11.002$ & [\ion{Ni}{3}] \\
    $        $ & $7.272$ & [\ion{Fe}{3}] & $      *\ *$ & $11.130$ & [\ion{Ni}{4}] \\
    $    *\ *$ & $7.349$ & [\ion{Ni}{3}] & $      *\ *$ & $11.167$ & [\ion{Co}{2}] \\
    $ *\ *\ *$ & $7.507$ & [\ion{Ni}{1}] & $      *\ *$ & $11.307$ & [\ion{Ni}{1}] \\
    $        $ & $7.773$ & [\ion{Co}{1}] & $*\ *\ *\ *$ & $11.888$ & [\ion{Co}{3}] \\
    $ *\ *\ *$ & $7.791$ & [\ion{Fe}{3}] & $         *$ & $11.978$ & [\ion{Fe}{3}] \\ 
    $       *$ & $8.044$ & [\ion{Co}{2}] & $      *\ *$ & $12.001$ & [\ion{Ni}{1}] \\ 
    $       *$ & $8.063$ & [\ion{Ni}{2}] & $      *\ *$ & $12.255$ & [\ion{Co}{1}] \\ 
    $        $ & $8.114$ & [\ion{Co}{1}] & $         *$ & $12.261$ & [\ion{Mn}{2}] \\ 
    $    *\ *$ & $8.211$ & [\ion{Fe}{3}] & $   *\ *\ *$ & $12.642$ & [\ion{Fe}{2}] \\ 
    $        $ & $8.282$ & [\ion{Ni}{1}] & $      *\ *$ & $12.681$ & [\ion{Co}{3}] \\ 
    $ *\ *\ *$ & $8.405$ & [\ion{Ni}{4}] & $   *\ *\ *$ & $12.729$ & [\ion{Ni}{2}] \\ 
    $    *\ *$ & $8.489$ & [\ion{Co}{3}] & $          $ & $13.058$ & [\ion{Co}{1}] \\ 
    \enddata
    \tablecomments{For each transition, the markers correspond to dominant
    ({$*\ *\ *\ *$}), strong ($*\ *\ *$), moderate ($*\ *$), weak ($*$), and scarcely 
    detectable ($~$) on top of the quasi-continuum formed by a large number 
    of lines. The relative strength S is estimated by the integral over the 
    envelope, $\int{ A_{ij}  n_j \,dV }$ where $n_j$ is the particle density 
    of the upper level. The list is based on the simulations described in 
    \autoref{sec:Mod}.}
\end{deluxetable}

\subsection{The $6.0-8.0$~\mic Region}

The $6.0-8.0$~\mic region is dominated by emission lines of 
stable Ni, the most prominent of which is a blend of 
[\ion{Ni}{3}]~$7.349$~\mic and [\ion{Ni}{1}]~$7.507$~\mic 
that defines the red edge of the feature. The blue edge of 
this peak is blended with several other weaker lines, creating 
a series of shoulders, extending from $\sim 6.5$~\mic to 
$\sim 7.2$~\mic. Moving from red to blue, these shoulders are 
comprised of [\ion{Ar}{2}]~$6.985$~\mic, [\ion{Ni}{2}]~$6.920$~\mic, 
and [\ion{Ni}{2}]~$6.636$~\mic. Finally, there is a small bump 
associated with a combination of [\ion{Co}{2}]~$6.214$~\mic, 
[\ion{Co}{1}]~$6.273$~\mic, and [\ion{Co}{2}]~$6.274$~\mic.

\subsection{The $8.0-9.5$~\mic Region}

The $8.0-9.5$~\mic region is dominated by two features whose edges 
are blended with other weaker lines at $\sim8.7$~\mic. The bluer 
of the two is due to the emission of [\ion{Ni}{4}]~$8.405$~\mic, while 
the redder feature is dominated by the [\ion{Ar}{3}] $8.991$~\mic 
line. The [\ion{Ar}{3}] line shows a distinct flat-topped profile,
which increases in flux moving from blue to red. Tilted flat-topped 
profiles are connected to both an ion's velocity distribution in the 
ejecta and the viewing angle of the explosion (see \autoref{sec:flat-top} 
and \autoref{sec:incangle}, also \citealt{Hoeflich_2021_20qxp}).
Small contributions from the weak [\ion{Fe}{2}]~$8.733$~\mic and 
[\ion{Ni}{4}]~$8.945$~\mic lines may also add to the observed flux
at the 10\% level.

\subsection{The $9.5-11.5$~\mic Region}

The $9.5-11.5$~\mic region shows a structure reminiscent of the 
$6.0-8.0$~\mic region; with one dominant blended feature and a series 
of smaller bumps and shoulders blended into the wings. The strongest 
peak arises from a blend of [\ion{Co}{2}]~$10.523$~\mic and 
[\ion{Ni}{2}]~$10.682$~\mic. A blend of [\ion{Fe}{2}]~$10.189$~\mic
and [\ion{Fe}{3}]~$10.203$~\mic forms a shoulder that is partially
blended into the blue wing of the [\ion{Co}{2}]$+$[\ion{Ni}{2}] blend.
Blended into the red wing is a series of three other weaker features.
The first feature, centered near $\sim10.85$~\mic is not associated with
any strong lines in our model. The next feature in the series arises 
from the comparatively weak [\ion{Ni}{3}]~$11.002$~\mic line, while 
a blend of [\ion{Ni}{4}]~$11.130$~\mic and [\ion{Co}{2}]~$11.167$~\mic 
forms a shoulder on the red wing of the [{\ion{Ni}{3}}] line. Finally, 
there may be a small contribution to the red wing of the 
[\ion{Ni}{4}]$+$[\ion{Co}{2}] shoulder from [\ion{Ni}{1}]~$11.307$~\mic.

\subsection{The $11.5-13.0$~\mic Region}

The $11.5-13.0$~\mic region contains the only relatively isolated, 
un-blended feature in the MIR spectrum, the [\ion{Co}{3}]~$11.888$~\mic
resonance line which produces the strongest line in the entire MIR 
spectrum. Our model shows weak contributions from 
[\ion{Fe}{3}]~$11.978$~\mic and [\ion{Ni}{1}]~$12.001$~\mic, however 
they only produce $\sim$1\% of the flux and do not alter the line 
profile in a significant manner (again see \autoref{tab:mir_lines}). 
A small shoulder at the edge of the red wing of the [\ion{Co}{3}] 
line is attributable to [\ion{Co}{1}]~$12.255$~\mic. A series of 
peaks between $12.5-13.0$~\mic suggests the presence of multiple 
weak lines, however the low S/N in this region prevents us from 
unambiguously identifying any lines. We tentatively identify the 
first peak with [\ion{Fe}{2}]~$12.642$~\mic and 
[\ion{Co}{3}]~$12.681$~\mic, and the second peak with 
[\ion{Ni}{2}]~$12.729$~\mic. Our model shows no strong lines in the 
vicinity of the third and final peak in the series.

\section{Velocity Distributions and Line Profiles} \label{sec:Vel}

In this section, we discuss the velocity distributions and line 
profiles of three important species in the ejecta: Ar, Co, and Ni. 
In discussing these velocities and profiles we reiterate that the 
current wavelength calibration of the MIRI/LRS observations is
accurate to $0.05-0.02$~\mic, with lower errors at longer wavelengths.
This corresponds to an error on the order of $\sim500$~\kms in the
[\ion{Co}{3}]~$11.888$~\mic line, and $\sim1400$~\kms in the
[\ion{Ni}{3}]~$7.349$~\mic line. Future updates to the JWST 
pipeline calibration files may increase the precision of  
these results. 

\subsection{[\ion{Ar}{3}]~$8.991$~\mic} \label{sec:flat-top}

\begin{figure}[t]
    \centering
    \includegraphics[trim=0cm 1.5cm 0cm
      2cm,clip=True,width=0.45\textwidth]{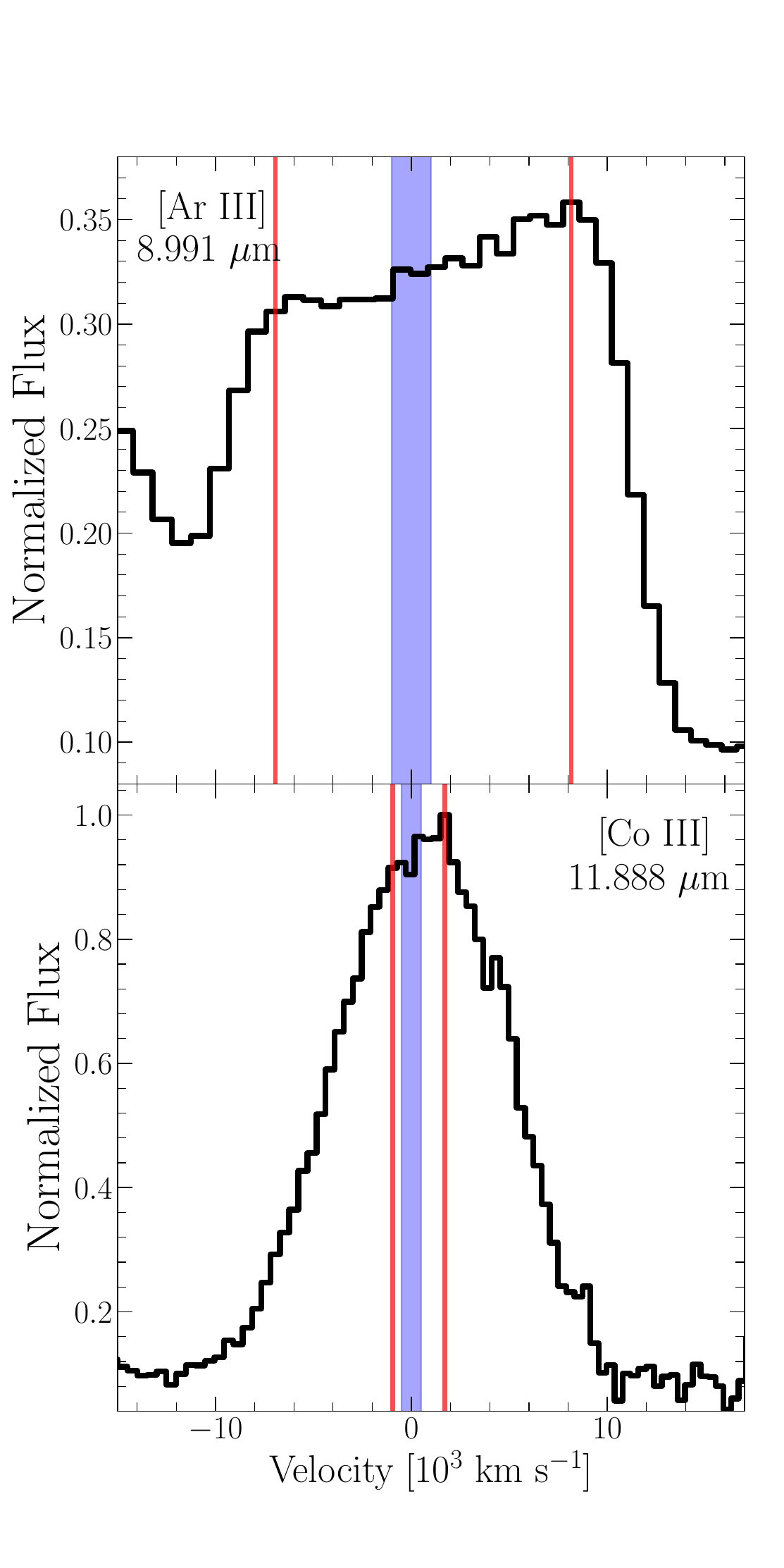} 
    \caption{Line profiles of the [\ion{Ar}{3}] (top) and [\ion{Co}{3}] 
    lines (bottom) in velocity space. The blue boxed region around $v = 0$~\kms 
    in the rest frame denote the 1$\sigma$ error in the rest wavelength
    for the given line. Red vertical lines mark the left and right edges 
    of the flat-tilted profiles in both panels.}  
    \label{fig:Arvel}
\end{figure}

Ar traces the transition region between incomplete oxygen burning 
and nuclear statistical equilibrium (NSE) in the ejecta; thereby providing 
details about the chemical distribution between the \Nifs and
Si-group layers. The [\ion{Ar}{3}]~8.991~\mic line profile is
plotted in \autoref{fig:Arvel} in velocity space. The profile is 
flat-topped with an increasing tilt from blue to red wavelengths,
which we refer as a ``flat-tilted'' profile hereafter. Flat-topped 
profiles are indicative of a central hole or void 
in the emitting material --- that is, a shell of line emitting material
\citep{Beals_1929_WR,Menzel_1929_WR,Struve_1931_Bstars}. For [\ion{Ar}{3}] the 
flat-top component of the feature starts at $\sim -7000$~\kms and extends 
to $\sim 8000$~\kms. The feature increases in flux by 10\% across the 
profile from the blue to red side, and the flat-topped component of the 
profile indicates that there is a central hole in the ejecta of 
$\sim \pm 8000$~\kms which does not contain Ar. This is because 
Ar is destroyed in high temperature regimes of the NSE where 
$T \geq 6 \times 10^9$~K, and there is a lack of strong mixing during
the explosion, consistent with explosion models of near~\Mch~WDs 
(see \autoref{sec:Mod} for details).  

\subsection{[\ion{Co}{3}]~$11.888$~\mic}

\begin{figure}[t]
  \centering
  \includegraphics[width=\columnwidth]{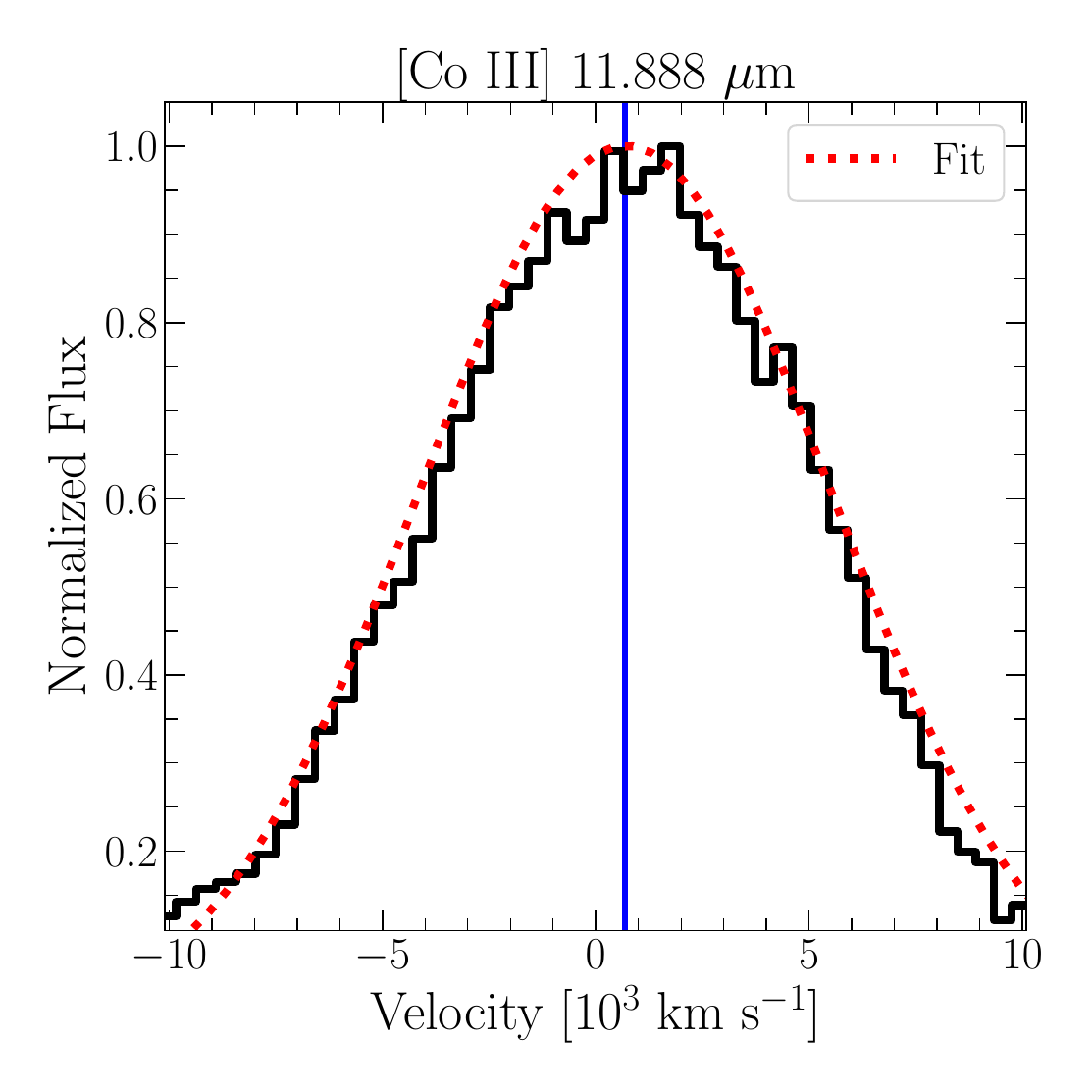}
  \caption{The [\ion{Co}{3}] 11.866 $\mu$m line compared to a Gaussian
    fit. The Gaussian peaks at $11.91 \pm 0.01$~\mic,
    $\sigma = 0.19 \pm 0.04$~\mic, which in velocity space corresponds
    to a peak at $740 \pm 200$~\kms and
    $\sigma = 4840 \pm 170$~\kms.}
  \label{fig:coiifit}
\end{figure}

SNe Ia are powered by the nuclear decay chain of \Nifs to $^{56}$Co to
$^{56}$Fe. Since the [\ion{Co}{3}]~$11.888$~\mic feature is a resonance 
line, most of the de-excitation and recombination of Co passes through 
this transition, making it
a direct tracer of the distribution and amount of \Nifs in the
ejecta. This feature covers a width of $\sim \pm 10000$~\kms.
If the shape of the line is assumed to be symmetric,
and thus well described by a Gaussian profile, it peaks at 
$740 \pm 200$~\kms with a FWHM of $4840 \pm 170$~\kms (see
\autoref{fig:coiifit}). Combining the error in the line-of-sight
velocity (recessional plus rotational; $\sim 60$~\kms) and the
estimated error in the wavelength calibration of $\sim 500$~\kms
with that of the fit error yields a total estimated error of $544$~\kms. 
The fact that this resonance line is not located at the kinematic center 
of the explosion indicates that the bulk of the \Nifs\ is off-center, at 
the $1.4\sigma$ level. The [\ion{Co}{3}]~$11.888$~\mic
profile also shows hints of a flat-tilted profile, peaking to 
the red at $\sim2000$~\kms (see \autoref{fig:Arvel}), although
the low resolution prevents a definitive identification of this
profile. Similarly, hints of this flat-tilted peak are also 
seen in the spectrum at the earlier epoch of SN~2021aefx 
\citep[see][]{Kwok2022}. If real, this flat-tilted profile may
extend from $\sim -1000$~\kms to $\sim 2000$~\kms. Similar to the 
[\ion{Ar}{3}]~$8.991$~\mic line, the flat-tilted profile of the [\ion{Co}{3}] 
feature may imply a central hole of \Nifs\ in the ejecta. This hole is 
smaller than that of Ar, and would only be a few thousand \kms
across \citep{Telesco_etal_2015,Diamond_etal_2015}. 
Note that unlike Ar, which is produced by nuclear burning that has
a steep temperature dependence leading to a sharp cutoff in velocity
extent and thus flat line profiles, electron capture is nearly
temperature independent, so its effects follow the density profile
leading to somewhat rounder line profiles.
The increase in flux across the [\ion{Co}{3}]~$11.888$~\mic profile is 10\%, 
the same as that in the [\ion{Ar}{3}]~$8.991$~\mic feature, implying
the distribution of the two elements are linked. Since Ar is produced 
at the edge of the \Nifs region (see \autoref{sec:Mod}), it is reasonable that Ar and Co have 
similar changes in flux across their profiles. In \autoref{sec:Mod}, we 
discuss the [\ion{Co}{3}]~$11.888$~\mic line profile in the context of 
off-center \Nifs distributions in the explosion. However, in order to 
confirm that the [\ion{Co}{3}]~$11.888$~\mic feature is truly 
asymmetrical and off-center, higher resolution spectra are required 
(such as those obtainable by the Medium Resolution Spectrograph (MRS)
of JWST/MIRI) and improved wavelength calibrations are also needed.

\subsection{Stable Ni} \label{sec:stableni}

\begin{figure*}
    \centering
    \includegraphics[width=\textwidth]{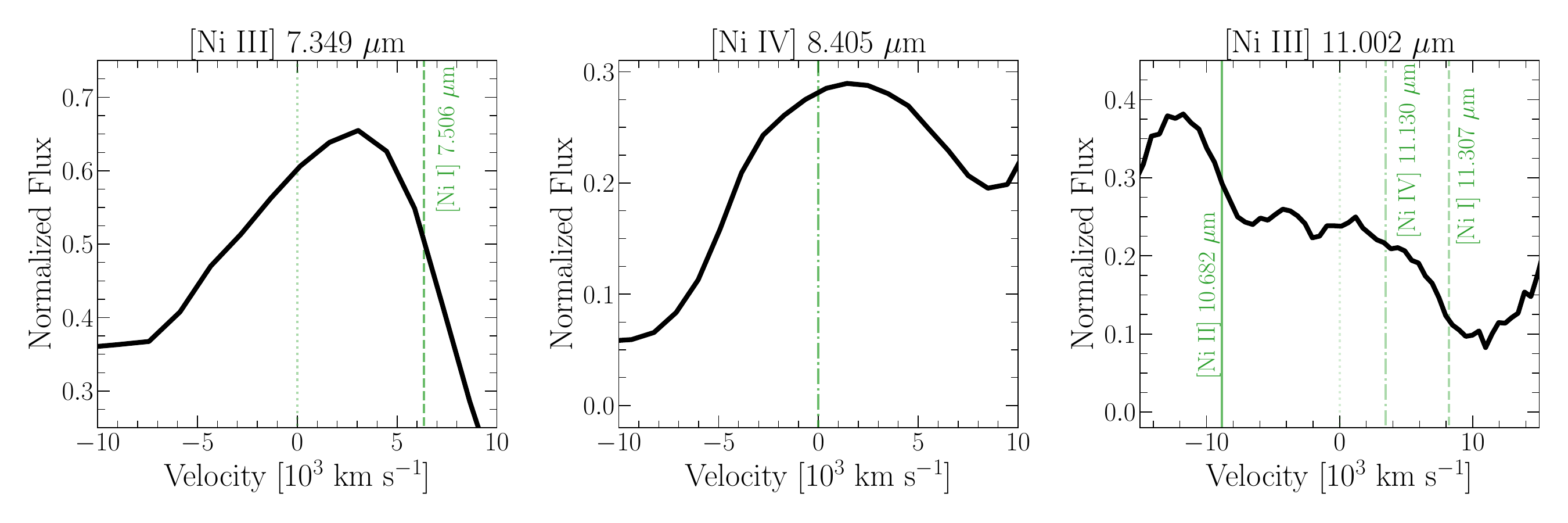}
    \caption{Velocity space profiles of the three spectral regions
    with prominent Ni lines. Vertical lines indicate the ionization
    and line strength as in \autoref{fig:IDS}. The left panel shows the 
    [\ion{Ni}{3}]~$7.349$~\mic region is contaminated by the 
    [\ion{Ni}{1}]~$7.506$~\mic feature. The right panel, centered
    on the [\ion{Ni}{3}]~$11.002$~\mic line, shows evidence for 
    multiple stable Ni lines contributing to the series of weak
    features and shoulders.}
    \label{fig:58nidistrubution} 
\end{figure*}

Multiple ionization states of Ni have forbidden emission lines
which occur in the MIR, making nebular phase MIR spectra an
invaluable resource for probing the explosion physics and 
corresponding nucleosynthesis of SNe~Ia. Since the \Nifs that powers
the early light curves of SNe~Ia has a half life of $6.1$~days, any
emission from Ni at these late phases comes from isotopes of stable 
Ni (e.g. $^{58}$Ni) and not from radioactive isotopes like \Nifs.  
\autoref{fig:58nidistrubution} presents three of these regions in 
velocity space. The left panel shows the [\ion{Ni}{3}]~$7.349$~\mic line,
which appears to be red-shifted in velocity space with an apparent 
maximum around $3000$~\kms. However, this feature is blended with 
 [\ion{Ni}{1}]~$7.506$~\mic, such that the velocity extent of 
[\ion{Ni}{3}]~$7.349$~\mic appears to be larger than its true 
distribution. The middle panel shows the [\ion{Ni}{3}]~$8.405$~\mic 
line profile in velocity space, while the right panel depicts the 
[\ion{Ni}{3}]~$11.002$~\mic feature within a much larger series of
blended lines. At higher resolution these blends, including the 
$11.002$~\mic line, are likely to be resolved.

\section{Numerical Modeling and Implications for Explosion Scenarios} \label{sec:Mod}

To explore the explosion physics of SN~2021aefx we turn to detailed
comparisons with NLTE radiation hydrodynamical models. The goals of 
these comparisons are: (1) to demonstrate that MIR spectral features 
and line profiles can be used as a critical tool to determine the 
explosion physics and progenitor scenario of SNe~Ia, (2) to show that 
JWST has opened up a new frontier in MIR SN science and that there is 
a need to test and calculate atomic models and processes, including 
cross sections to improve future models. Specifically, we address how 
the data allows us to measure the mass of the exploding WD, the chemical 
asymmetries in the initiation of the explosion, and small-scale mixing 
processes in the ejecta. When taken in total, these measurements allow 
us to determine the most likely explosion scenario of SN~2021aefx. 

As discussed in \autoref{sec:stableni} and shown by \citet{Kwok2022},
SN~2021aefx presents many spectral lines of stable Ni
(\autoref{fig:IDS}). This Ni requires high-density burning in the
ejecta, above 5 $\times$ 10$^8$~\gcm, which must originate from a
massive WD, making the explosion either a near-\Mch WD where the
explosion is triggered by compressional heating in the center of the
explosion, or the detonation of a high-mass, sub-\Mch larger than
1.15--1.2~\Msun
(\citealp{Hoeflich_Khokhlov_1996,1998ApJ...495..617H,Seitenzahl:2017},
 but see also \citealp{Blondin_Bravo_etal_2022}). Such massive WDs can 
only be produced via accretion \citep{2013sse..book.....K}. 
Therefore, we limit our comparisons to models within this region of
parameter space.

\subsection{Numerics}\label{sec:numerics}

The simulations employ modules of the HYDrodynamical
RAdiation~(HYDRA) code. HYDRA solves the time-dependent radiation
transport equation~(RTE) and positron transport
\citep{Penney_etal_2014}, including the rate equations that calculate
the nuclear reactions based on a network with 211 isotopes and
statistical equations for the atomic level populations, the equation
of state, the matter opacities, and the hydrodynamic evolution as
applied to SN~2020qxp \citep[][and references
therein]{Hoeflich_2021_20qxp,2021ApJ...923..210H}.  Detailed atomic
models and line lists are based on the database for bound-bound
transitions of \citet{vanHoof2018}\footnote{Version v3.00b3,
  \url{https://www.pa.uky.edu/~peter/newpage/}}, supplemented by
additional forbidden lines from \citet{Diamond_etal_2015} and
\citet{Telesco_etal_2015}. For details on modeling of nebular phase
spectra with HYDRA, see \citet{Hoeflich_2021_20qxp}, and for more
general discussions on modeling the nebular phase and downward
cascading of high-energy particles and photons by Monte Carlo, see also
\citet{Spencer_Fano_1954}, \citet{Axelrod_1980},
\citet{Kozma_Fransson_1992}, \citet{Fransson_1994_late}, 
\citet{2007Sci...315..825M}, \citet{2015MNRAS.450.2631M},
\citet{2015ApJ...814L...2F}, \citet{2017ApJ...845..176B}, \citet{Wilk2018},
\citet{Shingles2020}, and \citet{Wilk2020}. The models include
transitions for ionization stages I-IV of C, O, Ne, Mg, Si, S, Cl,
Ar, Ca, Sc, Ti, V, Cr, Mn, Fe, Co, and Ni. Though most of the
prominent features in the MIR are caused by forbidden lines, the
underlying quasi-continuum is formed by allowed lines in the inner
layers well above the critical density. 
At these phases, the iron-rich layers are still partially
optically thick at UV wavelengths, meaning the inclusion of permitted 
lines is important to fully characterize the ionization balance
via Rosseland cycles \citep{mihalas78sa}.

\subsection{A Delayed-Detonation Model for SN~2021aefx}\label{sec:model}

We compare SN~2021aefx to new simulations of off-center \Mch~mass 
explosion models, based upon the spherical model of the Model~25-series 
from \citet{Hoeflich_2017}, as it produces early light-curve properties 
and a maximum light luminosity very similar to those of SN~2021aefx.
These new simulations are parameterized explosion models, using a 
spherical delayed-detonation to constrain the global parameters of 
the explosion. Fine-tuning these models is not necessary to achieve 
the goals of this study as we focus on spectra rather than 
high-precision photometry. The model produces 
$\Delta m_{15}(V) = 0.68$~mag 
(for reference, $\Delta m_{15}(V) = 0.64 \pm 0.01$~mag for SN~2021aefx, 
which is within the error of the model), and $\sim$0.6~\Msun of \Nifs. 

The model originates from a C/O~WD with a main-sequence progenitor
mass of $5$~\Msun, solar metallicity, and a central density 
$\rho_c=1.1 \times 10^9$~\gcm. We adopt this $\rho_c$ due to the 
line width and shape of the [\ion{Co}{3}]~$11.888$~\mic line and 
due to the strength of the stable Ni lines in the MIR spectrum 
(see \autoref{sec:stableni}). In this model, burning starts as 
a deflagration front near the center and transitions to a detonation
\citep{1991A&A...245..114K}. The deflagration--detonation transition
is triggered when the density at the burning front drops below 
$2.5\times 10^7$~\gcm, when $\sim0.24$~\Msun of the material has 
been burned by the deflagration front, and is induced by the mixing 
of unburned fuel and hot ashes \citep{1991A&A...245..114K}. The model 
has a magnetic field of B(WD) $= 10^6$~G, which has been found in
magneto-hydrodynamical simulations, suggesting that turbulent magnetic 
fields are produced during the deflagration phase 
\citep{Diamond_etal_2018,2021ApJ...923..210H}. The basic model 
parameters are given in \autoref{tab:model_params}.

\begin{deluxetable}{lc}
  \tablecaption{Model 25 Parameters  \label{tab:model_params}}
  \tablehead{\colhead{Parameter} & \colhead{Value}}
  \startdata
    M$_{\rm ej}$ & $\sim$1.38 \Msun\\
    $\rho_c$ & $1.1 \times 10^9$ \gcm\\
    M$_{\rm tr}$ & $0.24$ \Msun\\
    M$_{\rm DDT}$ & $0.5$ \Msun\\
    B(WD) & $10^6$~G\\
  \enddata
\end{deluxetable}

\subsubsection{Off-center \Nifs and Abundance Distributions}

To investigate the line profiles and asymmetries, we consider the
Model 25-series which includes off-center DDTs.
For the construction of the off-center DDT we follow the 
description of \citet{1999ApJ...527L..97L} that has also been employed 
by \citet{2006NewAR..50..470H}, \citet{2015ApJ...804..140F}, 
\citet{Hoeflich_2021_20qxp}, and \citet{Hoeflich_etal_2023_19np}.  
The DDT is triggered at $M_\text{DDT}=0.5$~\Msun.
Note that due to the buoyancy of flame fronts in the explosion, the DDT
can be triggered at a different mass coordinate relative to the total
integrated mass of the deflagration burning. This leads to asymmetric
abundance distributions of all elements produced during the detonation
phase \citep[see Fig. 2 in][]{Hoeflich_2021_20qxp}. 

In principle, the use of multiple resolved line profiles allows us to
determine the value of $M_\text{DDT}$ as well as the viewing angle. In
the case of SN~2021aefx we use the two strongest features: 
[\ion{Co}{3}]~$11.888$~\mic and [\ion{Ar}{3}]~$8.991$~\mic. As shown 
in \autoref{fig:Arvel} we see a consistent tilt in the [\ion{Ar}{3}] 
and [\ion{Co}{3}] lines. We can determine the viewing angle from the 
tilt of these features. The value of $M_\text{DDT}$ 
determined here was also consistent with the
spectrophotmetric observations of the normal SN~Ia~2019np 
\citep{Hoeflich_etal_2023_19np}. Most normal SNe~Ia have 
very similar polarization properties \citep{Cikota_etal_2019}. 

\subsubsection{Overall Abundance Distribution}

\begin{figure*}[ht]
  \includegraphics[scale=0.7]{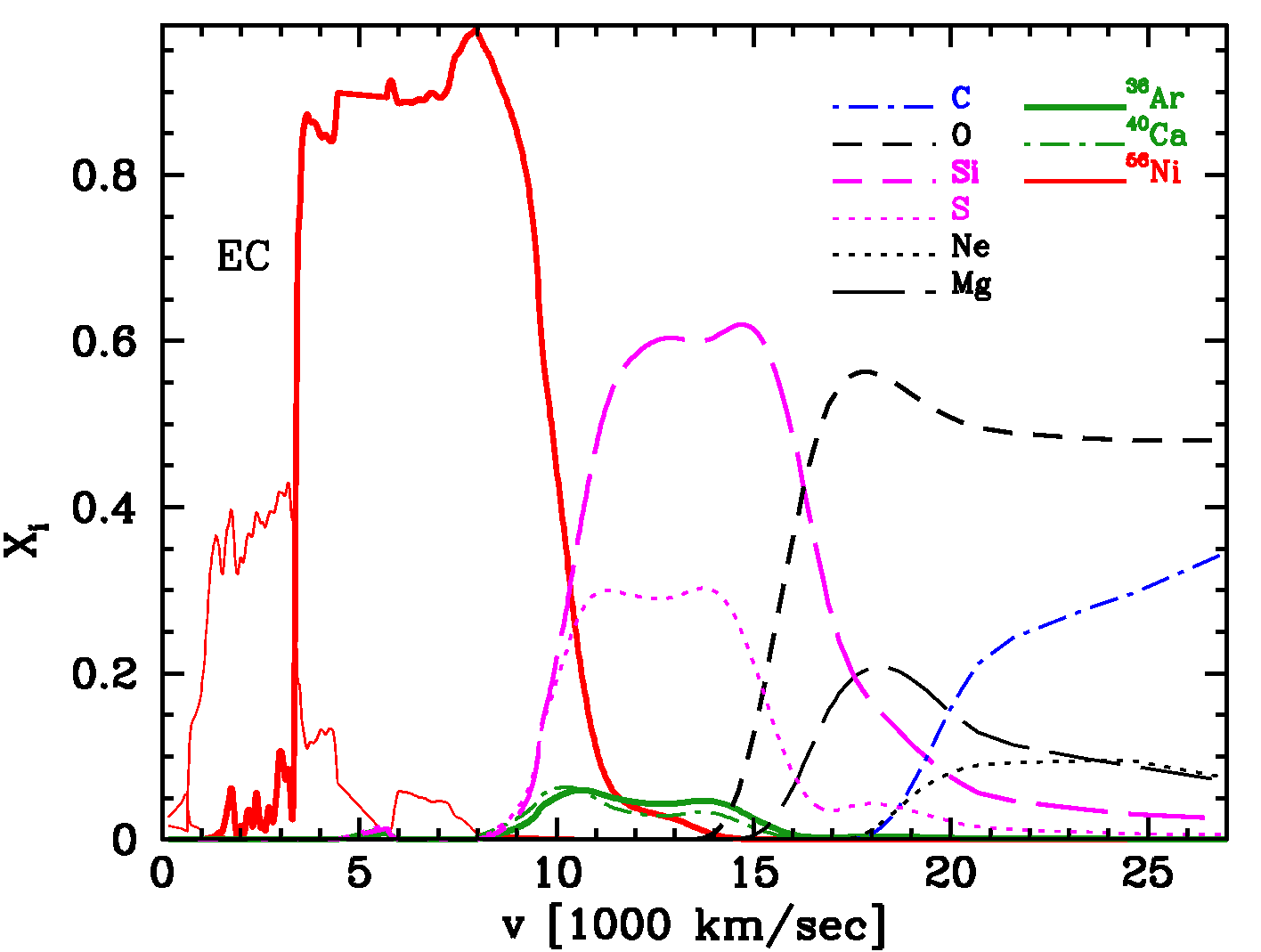}
  \includegraphics[scale=0.3]{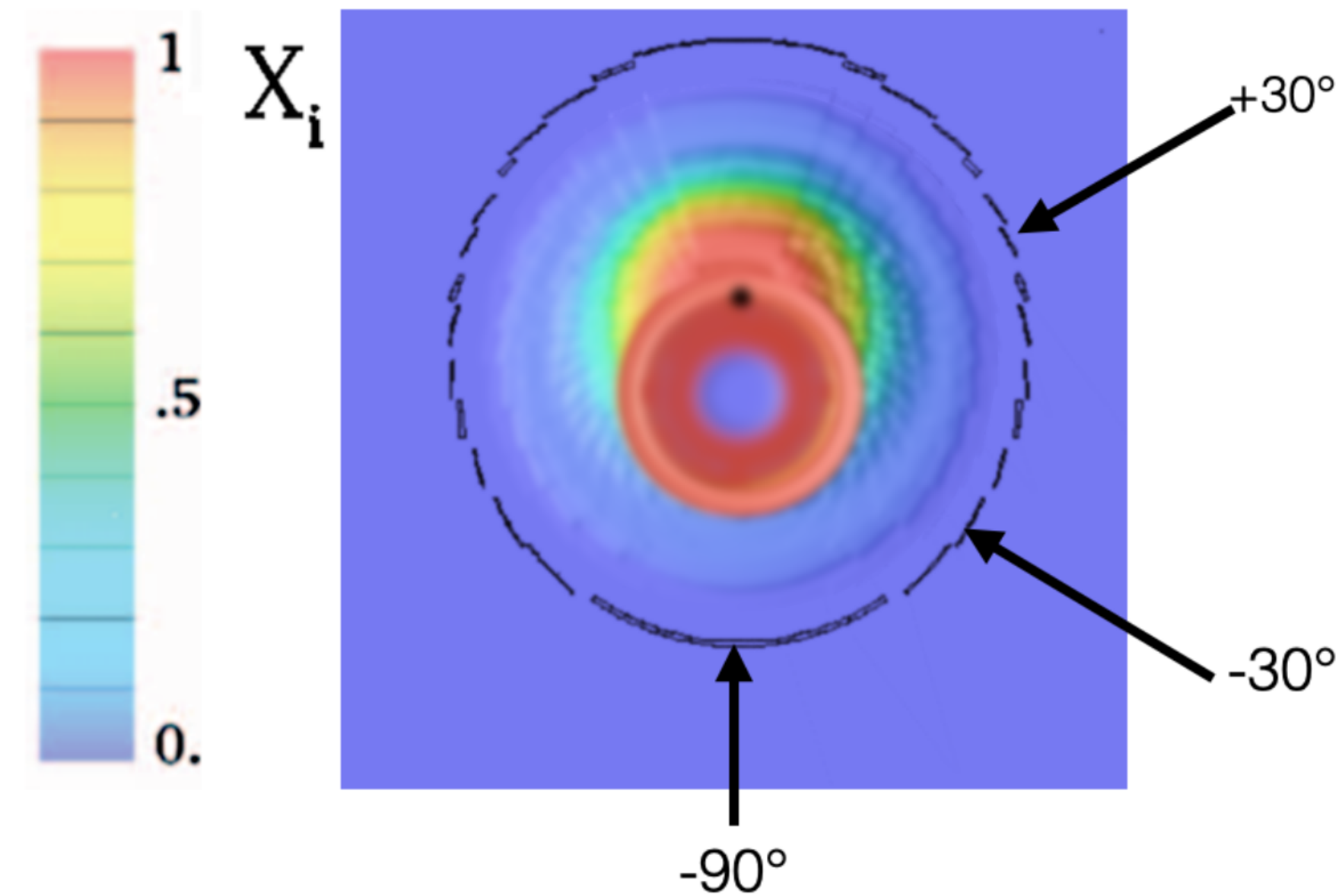}
  \caption
  {({\it Left}:) The chemical composition of our best fit model, Model~25
    from \citet{Hoeflich_etal_2017} and \citet{Hoeflich_etal_2023_19np}. 
    The model has a chemically stratified ejecta. EC elements (e.g. 
    $^{58}$Ni with M($^{58}$Ni) $\approx 5.9\times 10^{-2}$~\Msun) are
    located in the center of the ejecta, followed by \Nifs further
    out in velocity space. The Ar distribution goes between
    $8000$ -- $15000$~\kms and the lightest elements (e.g. O and C) are
    located in the outermost layers. For illustration, the thin red
    line at expansion velocities larger than $3200$~\kms shows the EC
    distribution after microscopic mixing applied (see text). ({\it Right}:)
    The distribution of the IGEs of the off-center
    delayed detonation Model~25 at a point (black dot). The bulk of
    the \Nifs is in a ring-like structure between $3000$ -- $9500$~\kms as
    well as a bulge produced at the point of the delayed detonation
    transition. Depending upon the viewing angle, differently shaped line 
    profiles will be produced in the [\ion{Co}{3}] feature. These profiles 
    are shown in \autoref{fig:coiiimodel_a}.}
  \label{fig:56Ni}
\end{figure*}

The angle-averaged abundance structure and the \Nifs distribution 
of Model~25 are shown in \autoref{fig:56Ni}. In the model, the region 
of high electron capture is spherical because we assume  central 
ignition, no fragmentation during the \Nifs decay over the first week 
after the explosion  \citep[e.g.][]{2015ApJ...804..140F}, and that 
Rayleigh-Taylor instabilities are largely suppressed by high magnetic 
fields \citep{Hristov_etal_2018}. The most notable results in the  
abundance distribution are: (1) $\sim 6 \times 10^{-2}$~\Msun of 
$^{58}$Ni is produced in the center of the ejecta, (2) the velocity 
extent of the central hole in \Nifs is $\sim 3200$~\kms, 
(3) the velocity extent of the \Nifs region 
produced in NSE ranges from $\sim 3200$ -- $10000$~\kms, and 
(4) the size of the shell of the Ar region covering a range of 
$\sim 8000$ -- $15000$~\kms. We note that simulating the point of the 
DDT in multi-dimensions does not lead to a strong rarefaction wave 
\citep{,Gamezo_etal_2005,2015ApJ...804..140F}
as seen in all spherical delayed-detonation models 
\citep{1991A&A...245..114K,Hoeflich_Gerardy_2002,Hoeflich_2021_20qxp}.

The off-center DDT at a point in an already-expanding medium results 
in a run-time effect which yields an asymmetric distribution of
burning products. The material closer to the DDT burns under 
higher density than the opposite side because the front reaches 
the corresponding layer $0.5- 1$~s later. The result is a bulge 
of all elements that undergo only Si and O burning including Ca, 
Ar, and \Nifs (see \autoref{fig:56Ni}). For a more complete 
depiction of this, see Fig. 7 of \citet{Fesen_etal_2007}. These 
asymmetries are aligned along the axis defined by the center and 
the DDT ignition point.

\subsection{Spectral Modeling}

\subsubsection{Determining the Inclination Angle} \label{sec:incangle}

We begin our discussion of Model 25's fit to the observations by 
illustrating its ability to determine the inclination angle of the 
explosion relative to our line of sight. Remember that to first order, 
the [\ion{Co}{3}] profile can be fit with a Gaussian of a half-width 
$\approx 4800$~\kms, emission wings ranging from $-10000$ -- $+10000$~\kms, 
and an offset from the rest wavelength of $+740$~\kms (see 
\autoref{sec:Vel} and \autoref{fig:coiifit}). This is 
consistent with the overall \Nifs distribution seen in the model 
(see \autoref{fig:56Ni}), but we note that assuming an emission feature 
is a Gaussian makes an implicit assumption about the underlying 
chemical distribution of an element within the ejecta, and should be 
used with caution.
As previously discussed, the host galaxy is seen face on and has a 
very small projected rotation ($65 \pm 60$~\kms), implying that host 
rotation plays a minor role in this offset. This leaves the peculiar 
motion of the progenitor system and the orbital velocity of the 
progenitor as remaining potential sources of this offset. However, if 
these were the dominant factors, one would expect a consistent velocity 
offset in all of the spectral lines, contrary to observations. 

\begin{figure*}[ht]
  \centering
  \includegraphics[width=\textwidth]{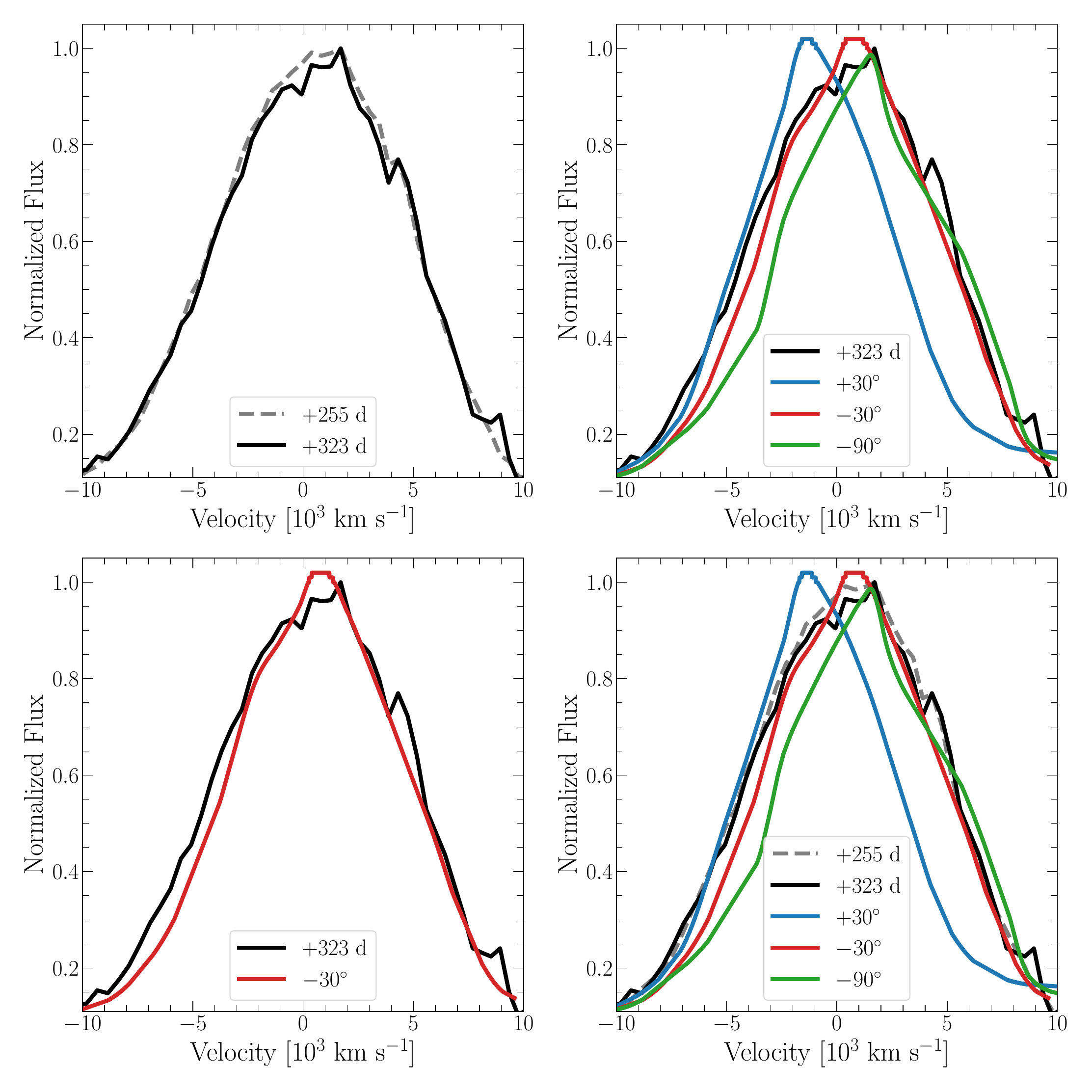}
  \caption{{\it Top Left}: Comparison of the [\ion{Co}{3}]~$11.888$~\mic
  line profile at $+255$ (dashed grey) and $+323$~days (solid black). 
  {\it Top Right}: Dependence of the [\ion{Co}{3}]~$11.888$~\mic line 
  profile as a function of inclination in comparison with the $+323$~day
  spectrum (solid black). Note that the profile and the red-shift of the 
  observed peak are consistent with the off-center DDT model seen 
  at $-30^\circ$ (bottom left) and the $-30^{\circ}$ model is also the best 
  fit to both observations (bottom right).}
    \label{fig:coiiimodel_a}
\end{figure*}

On the other hand, the observed flux in the [\ion{Co}{3}]~$11.888$~\mic
line center changes by $\sim 10\%$ across the peak of the feature (see 
 \autoref{sec:Vel}), consistent with expectations of flux arising 
from the asymmetric ejecta of an off-center DDT model when viewed from 
a specific angle \citep{Hoeflich_2021_20qxp}. As previously shown in 
\autoref{sec:Vel}, the [\ion{Co}{3}]~$11.888$~\mic feature appears to
show a flat-tilted profile, where the velocity extent of the central
tilted region corresponds to the region in velocity space of partial 
burning in quasi-statistical equilibrium~(QSE) 
($\approx 1800$~\kms across in the angle averaged spectrum). This 
flat-tilted profile is seen in both the $+255$ and $+323$ day JWST
spectra. In Model~25, the inner size of the electron capture region 
and the distribution of \Nifs produce different line profiles when 
viewed at different angles. Three specific viewing angles, $-90^\circ$, 
$-30^\circ$ and $+30^\circ$ are shown in \autoref{fig:coiiimodel_a}. 
From the bottom left panel, we see that the observations are well
matched by a viewing angle of $\approx -30^\circ$, including 
replicating the $\sim 10\%$ change in flux across the peak seen in 
the observations. While the high signal-to-noise (S/N $\approx 100$)
of both JWST spectra of SN~2021aefx suggests that the flat-tilted 
profile is real and significant, future planned observations with
JWST/MIRI MRS (JWST-GO-2114, PI: Ashall) will better resolve the
line profiles. 

\subsubsection{Overall MIR Spectra} \label{sec:obserallspec}

\begin{figure*}[ht]
  \includegraphics[width=\textwidth]{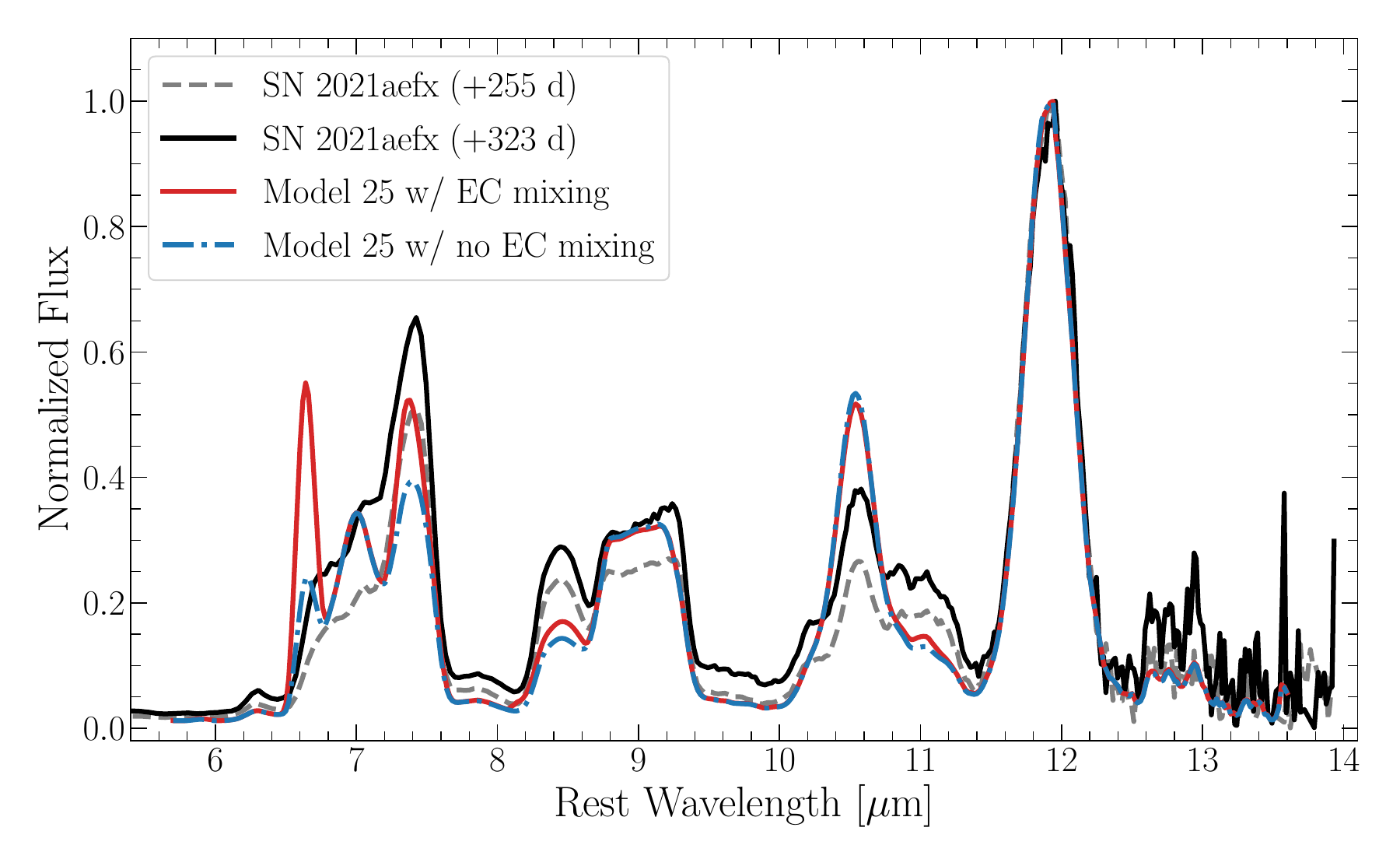}
  \caption{Comparison of the synthetic MIR spectrum of the off-center
    Model 25 seen from $-30^{\circ}$ without (blue) and with
    (red) mixing of the EC elements (see \autoref{fig:58nidistrubution}) 
    and the JWST/MIRI LRS spectrum of SN~2021aefx at $+255$ (dashed grey) 
    and $+323$ (solid black) days relative to $B$-band maximum. The 
    angle-averaged spectra would look similar, but they would show a 
    flat-topped rather than a flat-tilted  [\ion{Ar}{3}]~$8.991$~\mic feature. 
    Though [\ion{Ni}{2}] lines are present in both the synthetic spectra, 
    the sensitivity to microscopic mixing should be noted. In particular, 
    the [\ion{Ni}{2}] 6.6~\mic line shows a strong variation with mixing 
    (see text).} 
  \label{fig:specmod}
\end{figure*}

Having determined the inclination angle, we now compare our full model
spectrum to our observations, as seen in \autoref{fig:specmod}. Model
spectra are shown with and without mixing of electron capture elements
on the scale of the pressure scale height of the WD
\citep{Hoeflich_Stein_2002}. We examine microscopic mixing (\ie smaller 
than the mean free path of the positrons) in the
center of the explosion to constrain the position (e.g. central, 
off-center, or multi-spot) of the thermonuclear runaway ignition 
\citep{1996ApJ...471..903N,Hoeflich_Stein_2002,2004astro.ph..5162C,2005ApJ...632..443L,2007ApJ...660.1344R,2013ApJ...771...58M}.

The most dominant lines produced by
the models are shown in \autoref{tab:mir_lines}, and have been 
successfully identified in \autoref{fig:IDS}. For NIR lines outside 
the observed range, see \autoref{sec:nir_lines}. Identification of 
weaker lines within the spectrum will be possible after the 
acquisition of MIRI/MRS data.  
 
The model reproduces the observations overall, including all four 
regions of prominent spectral lines and does especially well in
reproducing the blends of the $6.0-8.0$~\mic and $8.0-9.5$~\mic
regions in addition to the [\ion{Co}{3}]~$11.888$~\mic line.
While the exact contribution of each ion may vary with the
underlying explosion model, the synthetic spectra have been obtained 
without further tuning and in general are in good agreement with 
the observations. The similarity between the mixed and unmixed model 
shows the stability of the synthetic spectral features. 

Most of the Ni features are in blends with other iron-group elements
of similar strengths. In light of uncertainties in the atomic models 
and cross sections, the photons at a given wavelength may couple to
elements other than Ni \citep[through fluorescence;][]{Morrison_Sartori_1966}. 
Thus, many of the line IDs of weaker features in low resolution 
spectra are model dependent. The
easiest way to separate the elements is by comparing the mixed and
unmixed models. In the unmixed models, the electron capture elements
are effectively shielded from non-thermal excitations from radioactive
decay, thus the electron capture features will be weaker. Features
dominated by Ni show variations in the region between $6.4$ -- $8.5$~\mic
and weak variation at longer wavelengths, for example at $10.5$~\mic.

One major effect of mixing can be seen in the $6.5$ -- $7.4$~\mic 
region. In the unmixed models, EC and radioactive elements are 
mostly separate, leading to weak spectral features. This due to 
the locally trapped positrons dominating the excitation in the EC 
region, with $\gamma$-rays responsible for $\sim10\%$ of the 
excited ions \citep{Penney_etal_2014}. In turn, microscopic mixing 
produces narrow features, for example, [\ion{Ni}{2}]
at $6.636$~\mic has a half-width of $\approx 4000$~\kms determined 
by the numerical resolution. In contrast, the observed broad
features suggest a central ignition of the WD.

The broad feature at $\sim9.0$~\mic dominated by Ar is a strong 
diagnostic of the point of the transition from the deflagration 
to the detonation. Ar is located in a shell with a large 
central hole because it is destroyed in moderately high density 
burning environments ($\rho \gtrsim 1-2 \times 10^7$~\gcm). Our 
model (\autoref{fig:coiiimodel_a}) is consistent with the estimates 
for the minimum Ar velocity, strongly suggesting a high mass
WD. The [\ion{Ar}{3}]~$8.991$~\mic feature shows the same slope as the 
[\ion{Co}{3}]~$11.888$~\mic, providing evidence of an off-center DDT.

Line profiles are strongly affected by the ionization balance.
Typically for normal SNe~Ia at this phase, iron group elements are
dominated by doubly ionized ions, and the ionization fraction
decreases with increasing density because the recombination rate
scales with the square of the density. Only in the center do we see
effects at the 10\% level of singly ionized iron group elements that 
produce a strong resonance [\ion{Ni}{1}] feature at $\approx 3$~\mic 
\citep{Kwok2022,Fisher_FSU_Phd_2022}. In the line forming regions,
the ionization balance hardly changes. Therefore, our results are 
insensitive to the differences in the ionization structure. More 
detailed discussions are given in \citet{Hoeflich_2021_20qxp}, 
\citet{Wilk2020}, and \autoref{sec:numerics}. 

In the synthetic spectrum, the feature at $\sim 11.9$~\mic agrees well 
with the observations, it is dominated by [\ion{Co}{3}]~$11.888$~\mic\ 
and has minor line blends of [\ion{Ni}{1}]~$12.001$~\mic and 
[\ion{Co}{1}]~$12.255$~\mic in the red wing at the 1\% level relative 
to the [\ion{Co}{3}] peak (see the line strengths given in \autoref{tab:mir_lines}).

However, our models tend to show features 
from singly ionized elements that are too strong by about $10-20$\% 
as can be seen from the ratio of the [\ion{Co}{3}] $11.888$~\mic 
and the [\ion{Co}{2}] $10.523$~\mic plus [\ion{Ni}{2}]~$10.682$~\mic blend. 
The discrepancies in the ionization balance are not unexpected, due to
uncertainties in the atomic data and the $\sim 2-5$\% accuracy of the 
flux calibration of the observed spectrum (see \autoref{sec:DR}).
Uncertain ionization and excitation by non-thermal leptons and 
uncertainties in the recombination rates lead to ionization balance 
uncertainties. Though we treat the cascades in energy within our 
Monte-Carlo scheme, missing and uncertain atomic levels are
likely responsible for some of the discrepancies
(\citealp{Wilk2018,Shingles2020}, and see Appendix A
of \citealp{Hoeflich_2021_20qxp}). 

Finally, we discuss other observables obtainable at a higher 
spectral resolution that may support our interpretation. Off-center
delayed-detonation models predict an offset between electron
capture elements (e.g., $^{58}$Ni) produced during the deflagration
phase and elements (e.g., $^{56}$Ni, Fe, Co, Si, S, Ar) synthesized 
during the detonation. The former are created in a subsonic deflagration
resulting in a slow pre-expansion phase of the WD with an almost
spherical density structure, whereas the latter are formed in a weak
detonation with a burning speed close to the speed of sound. This 
offset may be seen with higher-resolution spectra.

Note that the unresolved small flux variations near 11~\mic and in the 
wavelength range $12-14$~\mic seen in MIR spectra are at a 
$1\sigma $ level which, if confirmed by MRS spectra, have important 
implications. The computational results indicate that these small variations
signal the presence  
of a caustic structure in density and abundance of the inner electron-capture
core as has been observed by \citet{Fesen_etal_2007} and 
\citet{2015ApJ...804..140F}. At this epoch, positrons remain local for 
the high $B$-field required
\citep{Penney_etal_2014,2021ApJ...923..210H,2022ApJ...930..107M}.

\subsection{Alternative Explosion Scenarios} \label{sec:alt_scenario}

A detailed discussion of explosion scenarios and progenitor systems 
of SNe~Ia is beyond the scope of this work. For reviews, we refer to 
\citet{2017hsn..book.....A} and \citet{Hoeflich_2021_20qxp}. The 
total mass of the exploding WD is one of the parameters separating 
different explosion scenarios such as He-triggered detonations of 
sub-\Mch, dynamical mergers, violent mergers, collisions of two WDs 
in a triple system, and \Mch explosions. Dynamical mergers go 
through a loosely bound hydrostatic WD state and are unlikely to
synthesize EC elements because the peak density during merging is 
too low \citep{Benz_etal_1990,Garcia-Berro_etal_2017}. Collisions of 
two WDs may result in high density burning, 
high masses, and may produce EC elements; but also produce large 
polarization signatures and a $90^{\circ}$ flip in the polarization 
angle \citep{1995ApJ...440..821H,Bulla2016}, both of which are not 
observed in any SNe~Ia \citep{Cikota_etal_2019}. 
Similarly, violent mergers would be expected to yield high continuum 
polarization ($\sim1$\%) at $\sim 10$~days before the $B-$band 
maximum light \citep{Bulla2016}, which is also not seen in observations 
\citep{Yang2020,Patra2022}.

Therefore, we focus on sub-\Mch He-triggered detonations of C/O WDs 
as an alternative viable candidate to produce SN~2021aefx. For normal 
SNe~Ia sub-\Mch models have WD masses of $1$ -- $1.05$~\Msun and for 
bright SNe~Ia they have WD masses up to 1.1~\Msun 
\citep{Shen:2018,Blondin_Bravo_etal_2022}.

Stable Ni has been identified in observations 
 of several SN~Ia based upon heavily blended optical and NIR features 
 \citep[\eg][]{2018MNRAS.477.3567M,2020MNRAS.494.2809M}; however,  the Ni in 
 these features is never the dominant line. 
The 1.94~\mic [\ion{Ni}{2}] line may also be present in some SNe~Ia 
\citep{Friesen2014,2018A&A...619A.102D,Hoeflich_2021_20qxp}. 
However, this line is located at the edge of a telluric region, and 
with different reductions of the same data it has not been seen 
(see e.g. \citealt{2018A&A...619A.102D,Diamond_etal_2018}). 
Stable Ni has also been identified in the MIR spectra of some SNe~Ia
\citep{Gerardy2007,Telesco_etal_2015}, albeit at low S/N. However,
JWST spectra firmly establish the presence of stable Ni in SN~2021aefx
through multiple ionization states which are the dominant ions in 
their features; thus affirming that stable Ni exists in at least some 
SNe~Ia \citep{Kwok2022}. The presence of EC elements strongly suggest 
a high mass progenitor even within He-triggered detonations.

In the absence of detailed spectral models, we use as guide the 
$^{58}$Ni mass, $\approx 5.9 \times 10^{-2}$~\Msun obtained in our 
simulations. Although the overall abundance and density structures of 
various scenarios are similar, we note that the actual value of $^{58}$Ni 
depends on details of the structure and, possibly, microscopic mixing. 
For a \Mch progenitor, strong microscopic mixing can be excluded because 
it destroys the spectral fits (\autoref{fig:specmod}). 
We note that even larger uncertainties in M($^{58}$Ni) can be expected in 
1-zone models \citep{2020MNRAS.491.2902F} because they do not take into 
account variations in the chemical or density distributions in the inner 
layers and assume that each ion has a free density, meaning that the
ionization balance and level populations are not calculated self-consistently. 
Homogeneous abundances are unlike typical explosion models for SNe~Ia, and 
would require artificial mixing not to be expected in pure detonations.
For a sub-\Mch explosion, the EC and \Nifs\ region are produced in-situ, and
show boosted stable Ni line strengths but would not produce the asymmetric line 
[\ion{Ar}{3}] and [\ion{Co}{3}] profiles nor the correct width of [\ion{Co}{3}] 
feature. All three of these considerations point to a high 
likelihood of SN~2021aefx originating from a \Mch explosion.

Stable Ni is produced in NSE by shifting the ratio $Y_e$ from $\approx 0.5$ 
to a lower value either by electron-capture under high density burning or 
in a WD with super-solar metallicities, as a result of an initial high 
$^{22}$Ne abundance 
\citep[see, for example,][]{2000ApJ...536..934B,2003ApJ...590L..83T,2018SSRv..214...62T}. 
\citet{2021A&A...656A..94G} 
simulate a variety of He-detonation explosions at various metallicities, 
with various He-shell masses. They find that super-solar metallicities 
produce EC elements due to the decrease $Y_e$ from the presence of
neutron-rich $^{22}$Ne. \citet{2021A&A...656A..94G} find that WDs
with masses of $\sim1.1$~\Msun and primordial metallicity 3$Z_\odot$
produce $0.046$~\Msun of stable Ni, and the amount of
$^{58}$Ni sufficient to produce the strong Ni features observed in 
SN~2021aefx. However, since this result is due to the primordial 
metallicity of the WD, which reduced the $Y_e$ uniformly throughout 
the WD --- the full half-width of the [\ion{Co}{3}] line and of the 
Ni features should be comparable because both $^{56}$Ni and $^{58}$Ni 
are formed in the same NSE region and constant $Y_e$ results in a constant 
isotopic ratio. Future MIRI/MRS observations will be able to resolve 
the Ni lines and accurately measure its elemental distribution. In particular, 
in these sub-\Mch models the [\ion{Ni}{4}] should be broad as the high 
ionization stage will occur in the low-density, high-velocity region of the 
envelope because the recombination rates scale with the density squared
\citep{Osterbrock_Ferland_2006}. Moreover, the model that produces the large
$^{58}$Ni abundance requires of a  He shell of $0.02$~\Msun. Finally, if 
large stable Ni abundances are ubiquitous to all SNe~Ia, within the 
sub-\Mch paradigm it would require all of them to have super-solar metallicity, 
which is unlikely.

Based on recent 3D simulations for solar metallicities,
\citet{Boos_etal_2021} showed that a thin He-triggered detonation in a
$1.1$~\Msun C/O WD may produce $0.02$~\Msun of stable Ni,
an amount that is insufficient to explain the observed NIR features
from stable Ni \citep[see][]{Wilk2020}.\footnote{
  \citet{Blondin_Bravo_etal_2022} claim that sub-\Mch models with
  $M > 1$~\Msun can produce the NIR [\ion{Ni}{2}] lines at late
  times. They argue that the lack of [\ion{Ni}{2}] in the NIR at
  earlier times can be simply an ionization effect, with the mixing of
  radioactive products into the EC region dramatically reducing the
  strength of [\ion{Ni}{2}], thus, providing an option for the
  absence of the observed NIR [\ion{Ni}{2}] feature. This 
  explanation is, however, inconsistent with the fact that in 
  SN~2021aefx many ionization stages of Ni are observed.} 

For both sets of He-det simulations discussed above, the mass of the
outer He layers may also be inconsistent with recent 
limits from other early-time normal SNe~Ia spectra that show carbon 
in the outer $2-5 \times 10^{-3}$~\Msun 
\citep{Yang2020,Hoeflich_etal_2023_19np}. Furthermore, thin 
He-detonation models have nearly spherical \Nifs distributions  
\citep{Fesen_etal_2007,Hoeflich_etal_2023_19np}, which contradict 
the observation of SN~2021aefx.

Lacking advanced He-triggered detonation models, we focus on
spherical explosion models with sub-\Mch cores such as DET2
\citep{Hoeflich_Khokhlov_1996}. This model has a pure C/O WD without
a He surface layer. It originates from a WD whose mass is 1.2~\Msun, 
and produces a sufficient amount of EC material. Strict limits on the 
WD mass and the density of the central region can be obtained from both 
the [\ion{Ar}{3}]~$8.991$~\mic and [\ion{Co}{3}]~$11.888$~\mic features. 
In particular, a tight mass limit on the WD can be obtained via the width 
of the flat-top component of the [\ion{Ar}{3}]~$8.991$~\mic feature. The 
observed edge of the Ar profiles implies a central hole of Ar between 
$\sim 0 - 8000$~\kms (see \autoref{fig:Arvel}). However, DET2 produces an 
inner hole of Ar of $6000$~\kms. This is $2000$~\kms lower than 
observed. To produce a wider flat-topped Ar feature requires a 
higher mass WD. In a high mass model (such as a near \Mch explosion), 
the Ar hole extends further out in velocity space.

We note that, as an upper limit, detonating a WD with an ejecta mass
of 1.38~\Msun would mostly produce \Nifs and few QSE elements,
resulting in a SN that produced too much \Nifs, is too bright at
maximum light, and has spectra that are inconsistent with a typical SN
Ia \citep{1996ApJ...459..307H,Marquardt15}.

The velocity extent of the Ar hole would require a C/O WD of 
$\approx 1.24$~\Msun from HYDRA simulations. From stellar evolution 
\citep{Straniero:2016}, a maximum mass of a C/O core of 1.2~\Msun is
produced by stars with a main sequence mass of $\approx 8$~\Msun. For more 
massive progenitors, burning continues beyond He, resulting in O/Ne/Mg WDs 
and a core collapse SN \citep[see, for example][]{Woosley_Baron_1992}. 
Alternatively, O/Ne/Mg WDs that accrete material from a companion end 
their lives in accretion-induced collapse~(AIC) to a neutron star
\citep{Woosley_Baron_1992,Wasserburg:1996} because compression will
not reach temperatures in excess of $\approx 3 \times 10^9$~K needed
to trigger explosive O-burning. In the case of C/O detonation produced
via an external trigger (\citealp[\ie disruption by a black hole;][]{Rosswog_2009_BH}),
thermonuclear burning would result in a low-velocity explosion with
expansion velocities smaller by a factor of $2-3$ compared to typical
SNe~Ia \citep{Hoeflich_2017}. Thus, such C/O WDs can only be produced
via accretion over a long time \citep{2013sse..book.....K}.  This is,
however, inconsistent with the progenitor evolution channel commonly
assumed for He-shell detonations
\citep{Woosley_etal_1980,Nomoto_1982_p1,Hoeflich_Khokhlov_1996,
  Shen:2018}.

\section{Conclusion} \label{sec:conclus}

The successful launch of JWST heralds a new era in our understanding of 
the physics of thermonuclear supernovae. Late nebular phase MIR studies 
of SNe Ia are now possible thanks to JWST's impressive sensitivity, 
obtaining spectra with higher S/N and higher resolution than any prior 
MIR observatory capable of observing SNe~Ia.

Here, we present a JWST/MIRI LRS spectrum of SN 2021aefx at $+323$~days 
after maximum light obtained through JWST program GO-JWST-2114 
(P.I: C. Ashall). We show how a single spectrum can be used to 
extract previously unavailable information about SNe~Ia. We demonstrate 
how by combining JWST data with spectral models the nature of these 
important astrophysical objects can be determined. Below, we
highlight our most important results: 

\begin{itemize}

\item The observed spectrum of SN~2021aefx is linked to the physics 
of SNe~Ia through the construction of multi-dimensional radiation
hydrodynamical NLTE models.  We show that the spectrum and line
profiles can be understood within the context of a delayed-detonation
\Mch model that produced an asymmetric $^{56}$Ni distribution
originating from a WD with a central density of 
$\rho_{c} \approx 1.1 \times 10^9$~\gcm. 
Although it is at the brighter end of the 
distribution, the model for SN~2021aefx fits within the model series, 
which reproduces the light curves, the luminosity-width relation, 
and spectra of typical SNe~Ia (\autoref{sec:model}).

\item These models are used to identify the spectral lines which
  comprise the main features seen in the observed spectrum. The main 
  lines we identify include: [\ion{Co}{3}]~$11.888$~\mic,
  [\ion{Ar}{3}]~$8.991$~\mic, [\ion{Ni}{4}]~$8.945$~\mic,
  [\ion{Ni}{1}]~$7.507$~\mic, and [\ion{Ni}{3}]~$7.349$~\mic. Weaker
  identifiable blends include lines of: [\ion{Ar}{2}], [\ion{Fe}{2}],
  [\ion{Fe}{3}], [\ion{Co}{2}], and [\ion{Ni}{2}] (see \autoref{sec:Data}).

\item The presence of multiple Ni lines in the observed spectrum demonstrates
  that electron capture elements (e.g. $^{58}$Ni) are present in the
  inner region of SN~2021aefx. Significant amounts of these elements
  can only be produced by high-density burning (above $5 \times
  10^{8}$~\gcm). These densities are found in C/O WDs with masses above
   $\sim$~1.2~\Msun. Such massive WDs must be produced via accretion (see
  \autoref{sec:Vel}).

 \item We find evidence for no, or very limited, mixing on microscopic 
   scales between the electron capture elements and the \Nifs region in 
   the ejecta. In the context of  near \Mch models this suggests a central
   point of ignition (see \autoref{sec:obserallspec}). 
   Our simulations of SN~2021aefx suggest that 
   $\sim 0.06$~\Msun of stable Ni was produced in the explosion. In fact, 
   since the synthetic Ni lines appear slightly too weak 
   (\autoref{fig:specmod}), we may need slightly more stable Ni than our 
   simulations imply. The inclusion of mixing would not result in better
   agreement between the model and observations, as it 
   would alter the ionization balance of the Ni lines. 
   
\item Both the [\ion{Co}{3}]~$11.888$~\mic and
  [\ion{Ar}{3}]~$8.991$~\mic features show flat-tilted profiles, which
  vary by 10\% in flux across their peaks (see \autoref{sec:Vel}).
  These profiles are consistent with a central hole in the
  corresponding element distributions. The profiles also indicate
  that the explosion is seen at an inclination of $-30^\circ$ relative
  to the point of the deflagration to detonation transition (see
  \autoref{sec:incangle}).

\item We demonstrate how a flat-tilted profile can be used as a
  tool to determine the electron capture element and Ar
  distribution within the ejecta. The length of the flat-top component
  corresponds to the Doppler shift of the inner hole in the element
  distribution and the measured velocity extent corresponds to the average
  projected expansion velocity of the hole (see
  \autoref{sec:flat-top}).

 \item By combining information on both the strength and profiles of the
   Ar and stable Ni features, we show that SN~2021aefx was most likely
   produced from a C/O WD with a mass $> 1.2$~\Msun. This makes a
   He-detonation sub-\Mch explosion an unlikely candidate for this SN.  
   SN~2021aefx appears to be a normal SN~Ia with typical light curves 
   and spectra. We cannot rule out supersolar metallicity 
   in sub-\Mch WDs as an alternative option to produce EC elements in 
   SN~2021aefx, but such models would not produce a large 
   enough hole in the Ar region or the line profiles to reproduce the observations 
   of SN~2021aefx. Furthermore, we regard a He-detonation as
   unlikely, due to the fact that they produce spherical cores 
   (e.g. \Nifs\ distributions) which are not seen in SN~2021aefx, and 
   are also inconsistent with the carbon-rich surfaces commonly seen in 
   normal SNe~Ia (see \autoref{sec:alt_scenario}).

\end{itemize}

Although the data presented in this work have larger errors in wavelength 
calibration than anticipated, most aspects of the physical interpretation 
present here are insensitive to this error. For example, the off-center 
nature of the DDT is driven by the shape of the line profiles.
Moving forward, improved wavelength calibration from the JWST
pipeline, additional MIRI/LRS data (from program 2072; P.I: S. Jha), 
and future MIRI/MRS data (from program 2114; P.I: C. Ashall) will allow 
us to further constrain the physics of SN~2021aefx and other SNe~Ia, 
and can validate our interpretation. In particular, MIRI/MRS 
observations will improve the precision of the data by probing the SN 
ejecta to scales smaller than $\sim 100$~\kms which is essential. This MRS
data will also extend to longer wavelengths ($\sim20$~\mic) revealing
different lines and ions, as well as allowing us to identify weaker
features by resolving many of the blends seen in the LRS spectra. It 
will also open a new window to probe for smaller-scale effects such 
as mixing and positron transport within the ejecta at later times.

Overall, this work demonstrates the ability and potential that JWST
MIR spectral observations have to provide previously inaccessible
information to the scientific community. This new information will 
allow us to determine the progenitor scenario and explosion 
mechanism(s) of SNe~Ia. As the sample size of MIR spectra grows over 
the coming years we will be able to look for diversity within the SNe~Ia 
population.

\acknowledgements
JD, CA, PH, and EB acknowledge support by NASA grant
JWST-GO-02114.032-A. Support for program \#2114 was provided 
by NASA through a grant from the Space Telescope Science 
Institute, which is operated by the Association of Universities 
for Research in Astronomy, Inc., under NASA contract NAS 5-03127.
PH acknowledges support by the National
Science Foundation (NSF) through grant AST-1715133. EB acknowledges
support by NASA grant 80NSSC20K0538. 
This publication was made possible through the support of an LSSTC 
Catalyst Fellowship to KAB funded through Grant 62192 from the John 
Templeton Foundation to LSST Corporation. The opinions expressed in 
this publication are those of the author(s) and do not necessarily 
reflect the views of LSSTC or the John Templeton Foundation.
ID acknowledges partial support by the Spanish project 
PID2021-123110NB-100 financed by MCIN/AEI/10.13039/501100011033/FEDER/UE.
LG acknowledges financial support from the Spanish Ministerio de Ciencia 
e Innovaci\'on (MCIN), the Agencia Estatal de Investigaci\'on (AEI) 
10.13039/501100011033, and the European Social Fund (ESF) "Investing in your 
future" under the 2019 Ram\'on y Cajal program RYC2019-027683-I and the 
PID2020-115253GA-I00 HOSTFLOWS project, from Centro Superior de 
Investigaciones Cient\'ificas (CSIC) under the PIE project 20215AT016, and 
the program Unidad de Excelencia Mar\'ia de Maeztu CEX2020-001058-M.
The research of YY is supported through a Bengier-Winslow-Robertson Fellowship.
SWJ and LAK acknowledge support by NASA grant JWST-GO-02072.001 and 
NASA FINESST fellowship 80NSSC22K1599.
This work is based on observations made with the NASA/ESA/CSA James 
Webb Space Telescope. The data were obtained from the Mikulski Archive 
for Space Telescopes at the Space Telescope Science Institute, which 
is operated by the Association of Universities for Research in Astronomy, Inc., 
under NASA contract NAS 5-03127 for JWST. These observations are associated 
with program \#2114. The specific observations analyzed in this work can be 
accessed via \dataset[DOI: 10.17909/6fjc-sx91]{https://doi.org/10.17909/6fjc-sx91}.

\facilities{JWST (LRS/MIRI), MAST (JWST), The simulations presented here
were performed on the Beowulf system of the Astrophysics Group at Florida 
State University.}

\software{jwst (\citealp[ver. 1.8.1,][]{Bushouse2022_JWSTpipeline}),
          HYDRA \citep{Hoeflich2003,Hoeflich2009,Hoeflich_etal_2017}, 
          OpenDx (an open-sourced visualization package developed by IBM), 
          \texttt{SNooPy} \citep{Burns2011,Burns2014},
          Astropy \citep{astropy:2013, astropy:2018, astropy:2022},
          NumPy \citep{numpy2020}, SciPy \citep{SciPy2020}, 
          Matplotlib \citep{matplotlib}.}

\bibliographystyle{aasjournal}

\clearpage

\bibliography{ms}

\appendix

\section{Near-Infrared Line Identifications from Model 25} \label{sec:nir_lines}

\begin{deluxetable}{rcl|rcl|rcl|rcl|rcl}[h]
  \tablecaption{Near-Infrared Model Line Identifications \label{tab:nir_lines}}
  \tablehead{\colhead{\bf S} & \colhead{$\lambda$~[\mic]} & \colhead{Ion} 
    & \colhead{\bf S} & \colhead{$\lambda$~[\mic]} & \colhead{Ion}
    & \colhead{\bf S} & \colhead{$\lambda$~[\mic]} & \colhead{Ion}
    & \colhead{\bf S} & \colhead{$\lambda$~[\mic]} & \colhead{Ion}
    & \colhead{\bf S} & \colhead{$\lambda$~[\mic]} & \colhead{Ion}}
    \startdata
    $    *\ *$ & $2.211$ & [\ion{Fe}{2}] & $ *\ *\ *$ & $2.478$ & [\ion{Fe}{2}] & $       *$ & $2.935$ & [\ion{Co}{2}] & $        $ & $3.169$ & [\ion{Fe}{3}] & $ *\ *\ *$ & $4.076$ & [\ion{Fe}{2}] \\
    $ *\ *\ *$ & $2.219$ & [\ion{Fe}{3}] & $       *$ & $2.479$ & [\ion{Ni}{2}] & $       *$ & $2.954$ & [\ion{Co}{1}] & $        $ & $3.185$ & [\ion{Fe}{3}] & $       *$ & $4.077$ & [\ion{Fe}{3}] \\
    $       *$ & $2.219$ & [\ion{Fe}{3}] & $        $ & $2.481$ & [\ion{Co}{1}] & $    *\ *$ & $2.961$ & [\ion{Fe}{2}] & $       *$ & $3.187$ & [\ion{Co}{2}] & $    *\ *$ & $4.082$ & [\ion{Fe}{2}] \\
    $ *\ *\ *$ & $2.243$ & [\ion{Fe}{3}] & $        $ & $2.493$ & [\ion{Fe}{3}] & $        $ & $2.963$ & [\ion{Fe}{3}] & $ *\ *\ *$ & $3.230$ & [\ion{Fe}{3}] & $        $ & $4.108$ & [\ion{Co}{1}] \\
    $       *$ & $2.243$ & [\ion{Fe}{3}] & $        $ & $2.506$ & [\ion{Co}{1}] & $        $ & $2.965$ & [\ion{Fe}{3}] & $        $ & $3.230$ & [\ion{Fe}{3}] & $ *\ *\ *$ & $4.115$ & [\ion{Fe}{2}] \\
    $ *\ *\ *$ & $2.244$ & [\ion{Fe}{2}] & $    *\ *$ & $2.515$ & [\ion{Fe}{2}] & $        $ & $2.966$ & [\ion{Fe}{3}] & $       *$ & $3.239$ & [\ion{Co}{2}] & $       *$ & $4.307$ & [\ion{Co}{2}] \\
    $    *\ *$ & $2.257$ & [\ion{Fe}{2}] & $       *$ & $2.526$ & [\ion{Co}{1}] & $        $ & $2.987$ & [\ion{Fe}{3}] & $        $ & $3.242$ & [\ion{Fe}{3}] & $        $ & $4.340$ & [\ion{Co}{1}] \\
    $    *\ *$ & $2.267$ & [\ion{Fe}{2}] & $       *$ & $2.531$ & [\ion{Co}{2}] & $        $ & $3.006$ & [\ion{Co}{1}] & $        $ & $3.286$ & [\ion{Co}{2}] & $        $ & $4.357$ & [\ion{Fe}{3}] \\
    $    *\ *$ & $2.281$ & [\ion{Co}{3}] & $        $ & $2.570$ & [\ion{Fe}{3}] & $        $ & $3.012$ & [\ion{Co}{2}] & $        $ & $3.332$ & [\ion{Co}{1}] & $        $ & $4.357$ & [\ion{Fe}{3}] \\
    $       *$ & $2.282$ & [\ion{Co}{2}] & $        $ & $2.581$ & [\ion{Co}{1}] & $       *$ & $3.014$ & [\ion{Co}{3}] & $        $ & $3.353$ & [\ion{Co}{1}] & $       *$ & $4.410$ & [\ion{Co}{2}] \\
    $        $ & $2.284$ & [\ion{Co}{1}] & $       *$ & $2.601$ & [\ion{Co}{2}] & $       *$ & $3.014$ & [\ion{Co}{2}] & $       *$ & $3.394$ & [\ion{Ni}{3}] & $    *\ *$ & $4.435$ & [\ion{Fe}{2}] \\
    $        $ & $2.285$ & [\ion{Co}{1}] & $       *$ & $2.652$ & [\ion{Co}{1}] & $        $ & $3.017$ & [\ion{Fe}{3}] & $        $ & $3.471$ & [\ion{Co}{1}] & $       *$ & $4.520$ & [\ion{Ni}{1}] \\
    $        $ & $2.297$ & [\ion{Co}{1}] & $        $ & $2.686$ & [\ion{Co}{1}] & $        $ & $3.018$ & [\ion{Fe}{3}] & $       *$ & $3.492$ & [\ion{Co}{3}] & $ *\ *\ *$ & $4.608$ & [\ion{Fe}{2}] \\
    $    *\ *$ & $2.309$ & [\ion{Ni}{2}] & $       *$ & $2.692$ & [\ion{Co}{2}] & $       *$ & $3.031$ & [\ion{Co}{1}] & $        $ & $3.498$ & [\ion{Fe}{3}] & $       *$ & $4.672$ & [\ion{Fe}{2}] \\
    $        $ & $2.316$ & [\ion{Co}{1}] & $    *\ *$ & $2.717$ & [\ion{Fe}{3}] & $ *\ *\ *$ & $3.044$ & [\ion{Fe}{3}] & $       *$ & $3.630$ & [\ion{Co}{2}] & $       *$ & $4.788$ & [\ion{Ni}{1}] \\
    $       *$ & $2.335$ & [\ion{Ni}{2}] & $        $ & $2.717$ & [\ion{Fe}{3}] & $        $ & $3.044$ & [\ion{Fe}{3}] & $       *$ & $3.633$ & [\ion{Co}{1}] & $        $ & $4.860$ & [\ion{Fe}{3}] \\
    $        $ & $2.348$ & [\ion{Co}{1}] & $        $ & $2.726$ & [\ion{Co}{1}] & $        $ & $3.046$ & [\ion{Co}{1}] & $        $ & $3.647$ & [\ion{Co}{1}] & $ *\ *\ *$ & $4.889$ & [\ion{Fe}{2}] \\
    $ *\ *\ *$ & $2.349$ & [\ion{Fe}{3}] & $       *$ & $2.767$ & [\ion{Co}{2}] & $        $ & $3.061$ & [\ion{Co}{1}] & $        $ & $3.655$ & [\ion{Co}{1}] & $        $ & $5.054$ & [\ion{Co}{1}] \\
    $       *$ & $2.349$ & [\ion{Fe}{3}] & $        $ & $2.833$ & [\ion{Co}{1}] & $        $ & $3.063$ & [\ion{Fe}{3}] & $       *$ & $3.659$ & [\ion{Co}{2}] & $    *\ *$ & $5.062$ & [\ion{Fe}{2}] \\
    $       *$ & $2.361$ & [\ion{Ni}{2}] & $       *$ & $2.839$ & [\ion{Co}{2}] & $        $ & $3.085$ & [\ion{Fe}{3}] & $       *$ & $3.705$ & [\ion{Co}{2}] & $        $ & $5.164$ & [\ion{Co}{1}] \\
    $    *\ *$ & $2.370$ & [\ion{Ni}{2}] & $       *$ & $2.848$ & [\ion{Co}{2}] & $        $ & $3.085$ & [\ion{Fe}{3}] & $        $ & $3.738$ & [\ion{Co}{1}] & $       *$ & $5.180$ & [\ion{Co}{2}] \\
    $ *\ *\ *$ & $2.371$ & [\ion{Fe}{2}] & $       *$ & $2.871$ & [\ion{Co}{1}] & $        $ & $3.095$ & [\ion{Fe}{3}] & $        $ & $3.750$ & [\ion{Co}{1}] & $    *\ *$ & $5.187$ & [\ion{Ni}{2}] \\
    $        $ & $2.411$ & [\ion{Fe}{3}] & $ *\ *\ *$ & $2.874$ & [\ion{Fe}{3}] & $        $ & $3.097$ & [\ion{Fe}{3}] & $       *$ & $3.752$ & [\ion{Co}{2}] & $        $ & $5.211$ & [\ion{Co}{1}] \\
    $       *$ & $2.414$ & [\ion{Co}{1}] & $        $ & $2.874$ & [\ion{Fe}{3}] & $        $ & $3.100$ & [\ion{Fe}{3}] & $        $ & $3.771$ & [\ion{Co}{1}] & $ *\ *\ *$ & $5.340$ & [\ion{Fe}{2}] \\
    $        $ & $2.447$ & [\ion{Fe}{3}] & $       *$ & $2.889$ & [\ion{Co}{2}] & $       *$ & $3.100$ & [\ion{Co}{2}] & $       *$ & $3.802$ & [\ion{Ni}{3}] & $    *\ *$ & $5.674$ & [\ion{Fe}{2}] \\
    $       *$ & $2.453$ & [\ion{Fe}{3}] & $ *\ *\ *$ & $2.905$ & [\ion{Fe}{3}] & $        $ & $3.120$ & [\ion{Co}{1}] & $        $ & $3.823$ & [\ion{Co}{1}] & $    *\ *$ & $5.704$ & [\ion{Co}{2}] \\
    $        $ & $2.453$ & [\ion{Fe}{3}] & $        $ & $2.905$ & [\ion{Fe}{3}] & $ *\ *\ *$ & $3.120$ & [\ion{Ni}{1}] & $       *$ & $3.849$ & [\ion{Co}{2}] & $    *\ *$ & $5.893$ & [\ion{Ni}{1}] \\
    $    *\ *$ & $2.474$ & [\ion{Co}{3}] & $    *\ *$ & $2.911$ & [\ion{Ni}{2}] & $        $ & $3.129$ & [\ion{Fe}{3}] & $        $ & $3.877$ & [\ion{Co}{1}] & $       *$ & $5.940$ & [\ion{Co}{2}] \\
    $       *$ & $2.477$ & [\ion{Co}{2}] & $       *$ & $2.933$ & [\ion{Co}{2}] & $       *$ & $3.151$ & [\ion{Co}{2}] & $    *\ *$ & $3.952$ & [\ion{Ni}{1}] & $       *$ & $5.953$ & [\ion{Ni}{2}] \\
    \enddata
    \tablecomments{For each transition, the markers correspond to strong~($*\ *\ *$), 
    moderate~($*\ *$), weak~($*$), and scarcely detectable~(~) on top of the 
    quasi-continuum formed by a large number  of lines. The relative strength~S 
    is estimated by the integral over the envelope, $\int{ A_{ij} n_j \,dV }$ where 
    $n_j$ is the particle density of the upper level.}
\end{deluxetable}



\end{document}